\documentclass[pra,showpacs,groupedaddress,amssymb,twocolumn,notitlepage,nofootinbib,floatfix,superscriptaddress]{revtex4-1}
\usepackage{graphicx,amsmath,dsfont,physics}
\usepackage[T1]{fontenc}
\usepackage[dvipsnames]{xcolor}
\usepackage{subcaption}
\usepackage[english]{babel}
\usepackage{animate}
\usepackage{algorithm}
\usepackage{algpseudocode}
\usepackage{tabularx}
\usepackage{amsthm}
\usepackage{enumerate,mathtools}
\usepackage{float}
\usepackage{comment}
\usepackage[capitalize]{cleveref}
\usepackage{tikz}
\usetikzlibrary{arrows.meta, calc, decorations.pathmorphing, shapes, positioning,shapes.geometric} 
\usepackage{braket}
\makeatletter
\newcommand{\drawDetector}{
  \pgfpathellipse{\pgfpoint{0cm}{0cm}}{\pgfpoint{0.25 cm/4}{0}}{\pgfpoint{0}{0.25 cm/4*sin(15)}}
  \pgfusepath{stroke,fill}  

  \color{black}
  \pgfpathmoveto{\pgfpoint{0.25cm}{0cm}}
  \pgfpatharc{0}{180}{0.25cm}
  \pgfpatharc{180}{0}{0.25 cm and {0.25 cm*sin(15)}}
  \pgfusepath{stroke, fill}

  \pgfpathellipse{\pgfpoint{0cm}{0cm}}{\pgfpoint{0.25 cm}{0}}{\pgfpoint{0}{0.25 cm*sin(15)}}
  \pgfusepath{stroke}
}
\pgfdeclareshape{BSM_shape}{
  \anchor{center}{\pgfpoint{0cm}{0cm}}
  \anchor{north}{\pgfpoint{0cm}{sin(45) * 0.5 cm}}
  \anchor{north west}{\pgfpoint{-cos(45) * 0.8 cm}{sin(45) * 0.75 cm + 0.25 cm}}
  \anchor{north east}{\pgfpoint{cos(45) * 0.8 cm}{sin(45) * 0.75 cm + 0.25 cm}}
  \anchor{south}{\pgfpoint{0cm}{-sin(45) * 0.5 cm}}
  \anchor{south west}{\pgfpoint{-cos(45) * 0.25 cm}{-sin(45) * 0.25 cm}}
  \anchor{south east}{\pgfpoint{cos(45) * 0.25 cm}{-sin(45) * 0.25 cm}}
  
  \backgroundpath{
    \pgfscope
      \pgftransformshift{\pgfpoint{-cos(45) * 0.5 cm}{sin(45) * 0.5 cm}}
      \pgftransformrotate{45}
      \drawDetector
    \endpgfscope
    \pgfscope
      \pgftransformshift{\pgfpoint{cos(45) * 0.5 cm}{sin(45) * 0.5 cm}}
      \pgftransformrotate{-45}
      \drawDetector
    \endpgfscope

    \pgfpathmoveto{\pgfpoint{-cos(45) * 0.5 cm}{sin(45) * 0.5 cm}}
    \pgflineto{\pgfpoint{cos(45) * 0.25 cm}{-sin(45) * 0.25 cm}}
    \pgfpathmoveto{\pgfpoint{cos(45) * 0.5 cm}{sin(45) * 0.5 cm}}
    \pgflineto{\pgfpoint{-cos(45) * 0.25 cm}{-sin(45) * 0.25 cm}}
    \pgfusepath{stroke}

    \pgfsetstrokecolor{black}
    \pgfsetfillcolor{cyan!70!white}
    \pgfsetfillopacity{0.3}
    \pgfpathmoveto{\pgfpoint{0 cm}{sin(45)* 0.5 cm}}
    \pgflineto{\pgfpoint{cos(45)* 0.5 cm}{0 cm}}
    \pgflineto{\pgfpoint{0 cm}{-sin(45)* 0.5 cm}}
    \pgflineto{\pgfpoint{-cos(45)* 0.5 cm}{0 cm}}
    \pgfpathclose
    \pgflineto{\pgfpoint{0 cm}{-sin(45)* 0.5 cm}} 
    \pgfusepath{fill, stroke}
  }
}
\pgfkeys{
  /pgf/decoration/loops/radius A/.initial=0.25cm,
  /pgf/decoration/loops/radius B/.initial=0.2cm,
  /pgf/decoration/loops/extra loops/.initial=0,
  /pgf/decoration/loops/side/.initial=1,
}
\pgfdeclaredecoration{loop shape}{initial}{
  \state{initial}[width=0pt,next state=final]{
        \pgfmathsetlength{\pgf@xa}{\pgfkeysvalueof{/pgf/decoration/loops/radius A}}
        \pgfmathsetlength{\pgf@xb}{\pgfkeysvalueof{/pgf/decoration/loops/radius B}}
        \pgfmathsetmacro{\side}{\pgfkeysvalueof{/pgf/decoration/loops/side}}
        \pgfpathlineto{\pgfpoint{\pgfdecoratedpathlength-\pgf@xa}{0}}
        \pgfpatharc{-\side*90}{0}{\pgf@xa}
  }
  \state{final}{
        \pgfmathsetlength{\pgf@xa}{\pgfkeysvalueof{/pgf/decoration/loops/radius A}}
        \pgfmathsetlength{\pgf@xb}{\pgfkeysvalueof{/pgf/decoration/loops/radius B}}
        \pgfmathsetmacro{\side}{\pgfkeysvalueof{/pgf/decoration/loops/side}}  
        \pgfmathsetmacro{\numl}{\pgfkeysvalueof{/pgf/decoration/loops/extra loops}}
        \ifnum\numl>0
        \foreach \x in {1,...,\numl}{
            \pgfpatharc{0}{\side*360}{\pgf@xb}
            \pgflineto{\pgfpointadd{\pgfpoint{\pgfdecoratedpathlength}{\side*\pgf@xa}}{\x*\side*\pgfpoint{0}{0.025 cm}}}    
        }
        \fi
        \pgfpatharc{0}{\side*180}{\pgf@xb}
        \pgflineto{\pgfpoint{\pgfdecoratedpathlength-2*\pgf@xb}{\side*\pgf@xa}}
        \pgfpatharc{180}{180+\side*90}{\pgf@xa}
        \pgfpathlineto{\pgfpointdecoratedpathlast}
  }
}

\newtheorem{semiclassicallimit}{Semiclassical Limit}

\newcounter{protocol}

\usepackage{color}
\usepackage{amsthm}
\usepackage{amsmath,amssymb,amsbsy}

\usepackage{xcolor}
\pagecolor{white}
\usepackage[english]{babel}

\usepackage{calrsfs}

\newcommand{\ones}{\boldsymbol{1}}
\newcommand{\zeros}{\boldsymbol{0}}

\renewcommand{\Tr}{\mathrm{Tr}}

\newcommand{\expect}[1]{\left\langle#1\right\rangle}
\newcommand{\eea}{\end{eqnarray}}
\newcommand{\bea}{\begin{eqnarray}}
\newcommand{\ee}{\end{equation}}
\newcommand{\be}{\begin{equation}}

    \newcommand{\idh}{\hat{\mathds{1}}}

\tikzset{%
    light/.style = {
        color = red!70!black
    },
    Spin/.pic = {
        \draw[-{Latex[round, length=4pt]}, color = red!50!black, thick] (0, -0.2) -- (0, 0.2);
        \fill[color = red!50!black] (0, 0) circle (0.05);
    },
    BSM/.style = {draw, light, shape = BSM_shape, scale = 0.6},
    Det/.pic={
        \draw[fill=black] (0.25,0) arc (0:180:0.25) arc  (180:0:0.25 and sin{15}*0.25);
        \draw (0,0) circle [x radius = 0.25, y radius= 0.25*sin{15}];
        \filldraw[light] (0,0) circle [x radius = 0.25/4, y radius = 0.25*sin{15}/4] ;
    },
    pics/Pulse/.style={code ={
        \filldraw[domain=-0.4:0.4,smooth, light, scale = 0.35, fill = red!70!black!60!white] plot  (\x,{#1*exp(- \x*\x/0.03)});
    }
    },
    pics/laser/.style = {code ={
      \node[rectangle, minimum width = 1 cm, minimum height = 0.25 cm, draw] (las) at (-0.2,0) {};
      \fill ($(las.north)!0.7!(las.north east)$) -- ($(las.south)!0.7!(las.south east)$) -- ($(las.south)!0.5!(las.south east)$) to[bend right] ($(las.north)!0.5!(las.north east)$) -- cycle;
      \fill (las.south west) -- (las.north west) -- ($(las.north west)!0.2!(las.north)$) to[bend right] ($(las.south west)!0.2!(las.south)$) -- cycle;
      \fill[ color = red!30!white] (las.north east) -- ++(0.5,-0.1) -- ++ (0,-0.05)  -- (las.south east) -- cycle;
      \foreach \i in {-90, -45, ..., 90}{
        \draw[ thick, color = red!30!white] ($(las.east) + (0.5,0)+ (\i:0.05)$) --++ (\i:0.1);
      }}
    },
    client/.style={
        regular polygon,
        regular polygon sides=6,
        minimum size = 0.75 cm,
        thick,
        draw = black
    },
    server/.style = {
        circle,
        minimum size = 0.75 cm,
        thick,
        draw = black,
    }
}

\begin{document}
\title{Quantum Strategies to Overcome Classical Multiplexing Limits}

\author{Tzula B. Propp}
\affiliation{QuTech, Kavli Institute of Nanoscience\\
TUDelft, Lorentzweg 1, 2628 CJ Delft, Netherlands}
\affiliation{Leiden Institute of Physics\\ Leiden University, 2333 CA Leiden, The Netherlands}
\author{Jeroen Grimbergen}
\affiliation{QuTech, Kavli Institute of Nanoscience\\
TUDelft, Lorentzweg 1, 2628 CJ Delft, Netherlands}
\author{Emil R. Hellebek}
\affiliation{Center for Hybrid Quantum Networks (Hy-Q), Niels Bohr Institute, University of Copenhagen, Jagtvej 155A, Copenhagen DK-2200, Denmark}
\author{Junior R. Gonzales-Ureta}
\affiliation{Q*Bird, Delftechpark 1, 2628 XJ, Delft, The Netherlands}
\author{Janice van Dam}
\affiliation{QuTech, Kavli Institute of Nanoscience\\
TUDelft, Lorentzweg 1, 2628 CJ Delft, Netherlands}
\author{Joshua A. Slater}
\affiliation{Q*Bird, Delftechpark 1, 2628 XJ, Delft, The Netherlands}
\author{Anders S. Sørensen}
\affiliation{Center for Hybrid Quantum Networks (Hy-Q), Niels Bohr Institute, University of Copenhagen, Jagtvej 155A, Copenhagen DK-2200, Denmark}
\author{Stephanie D. C. Wehner}
\affiliation{QuTech, Kavli Institute of Nanoscience\\
TUDelft, Lorentzweg 1, 2628 CJ Delft, Netherlands}

\begin{abstract}

Near-term quantum networks face a bottleneck due to low quantum communication rates. This degrades performance both by lowering operating speeds and increasing qubit storage time in noisy memories, making some quantum internet applications infeasible. One way to circumvent this bottleneck is multiplexing: combining multiple signals into a single signal to improve the overall rate. Standard multiplexing techniques are classical in that they do not make use of coherence between quantum channels nor account for decoherence rates that vary during a protocol's execution. In this paper, we first derive semiclassical limits to multiplexing for many-qubit protocols, and then introduce two techniques: single click quantum multiplexing and multi-server multiplexing. These can enable beyond-classical multiplexing advantages. We illustrate these techniques through three example applications: 1) entanglement generation between two assymetric quantum network nodes (i.e., repeaters or quantum servers with inequal memories), 2) remote state preparation between many end user devices and a single quantum node, and 3) remote state preparation between one end user device and many internetworked quantum nodes. By utilizing many noisy internetworked quantum devices instead of fewer low-noise devices, our multiplexing strategies enable new paths towards achieving high-speed many-qubit quantum network applications.

\end{abstract} 

\maketitle
\vspace{-1em}

\section{Introduction}
\vspace{-1em}




Multiplexing is a canonical technique to multiplicatively improve channel capacity \cite{Nyquist1928,Shannon1949}; by laying down $M$ times as much fiber or by multiplexing signals into $M$ co-propagating frequencies, $M$ times as many classical bits may be transmitted per second \cite{Stallings2006}, improving the execution rate of the overall protocol or application. In quantum science, classical multiplexing of flying qubits (i.e. photonic qubit encodings \cite{Cirac1997,Duan2001,gheri1998photon}) allows independent quantum signals to co-propagate in a channel e.g. by utilizing separate time-bins \cite{Mueller2024} (Fig. 1a), spatial modes \cite{Ortega2024}, or frequencies \cite{Alshowkan2022}. This improves the rate of transmitted quantum information along a quantum network \cite{Caves1994,Lloyd1997,Collins2007,Bacco2019,Iesta2024}, as well as enables new device capabilities (e.g. number resolution in single photon detectors \cite{Nehra2020}).

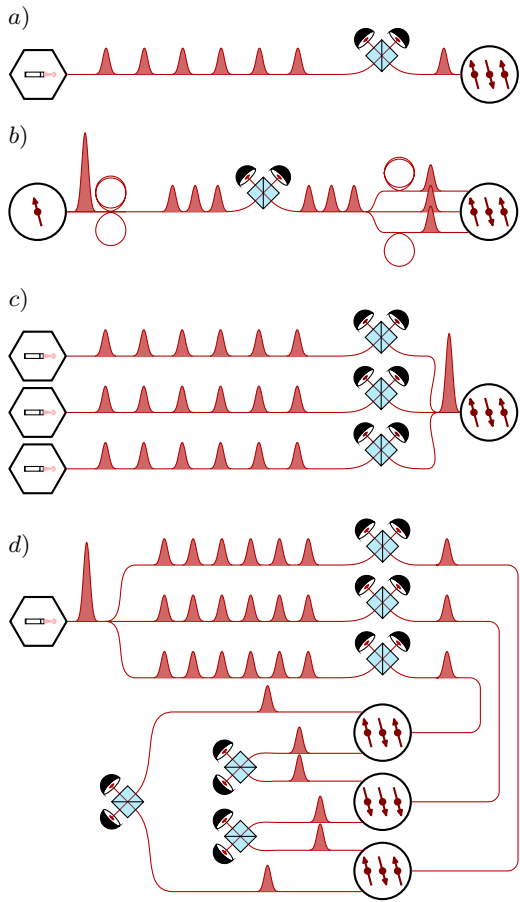
\begin{figure}[t]
    \centering
    \begin{tikzpicture}
        \node[node font = \small] (label) at (-1/4,0.75) {$a)$};
        \node[client] (cl) at (0,0) {};
        \pic[scale = 0.25, transform shape] at (cl) {laser};
        \node[server] (QBB) at (6,0) {};
        \foreach \i in {-1,0,1}{
            \draw ($(QBB) + (\i*0.2,0)$) pic [rotate = 195+\i*180, transform shape] {Spin};
        }
        \node[BSM] (BSM) at ($(cl.east)!0.8!(QBB.west) +(0,0.25)$) {};
        \draw[light] (cl.east) -- ($(QBB.west)!(BSM.south west)!(cl.east) - (0.5,0)$) foreach \j in {1, ..., 6}{
                pic [pos = \j/7] {Pulse=1}
            } to[out = 0, in = -135] (BSM.south west);
        \draw[light] (QBB.west) -- ($(QBB.west)!(BSM.south east)!(cl.east) - (-0.5,0)$) 
            pic [pos = 1/2] {Pulse=1}
            to[out = 180, in = -45] (BSM.south east);
    \end{tikzpicture}

    \vspace{1 ex}
    \begin{tikzpicture}
        \node[node font = \small] (label) at (-1/4,1) {$b)$};
        \node[server] (QBA) at (0,0) {};
        \draw (QBA) pic [rotate = 15, transform shape] {Spin};
        \node[server] (QBB) at (6,0) {};
        \foreach \i in {-1,0,1}{
            \draw ($(QBB) + (\i*0.2,0)$) pic [rotate = 195+\i*180, transform shape] {Spin};
        }
        \node[BSM] (BSM) at ($(QBA.east)!0.5!(QBB.west) +(0,0.25)$) {};
        \coordinate (BSMbw) at ($(QBA.east)!(BSM.south west)!(QBB.west)$);
        \coordinate (BSMbe) at ($(QBA.east)!(BSM.south east)!(QBB.west)$);
        \coordinate (split) at ($(QBA.east)!0.15!(QBB.west)$);
        \coordinate (together) at ($(QBB.west)!0.5!(BSMbe)$) ;
        \draw[light] decorate [decoration={loop shape, loops/side = -1}]  {(QBA.east) -- (split)};
        \pic at ($(QBA.east)!0.5!(split)-(0.15,0)$) {Pulse=3};
        \draw[light] 
            decorate [decoration={loop shape, loops/extra loops = 1}]  
            {(QBA.east) -- (split)} -- ($(BSMbw) - (0.5,0) $)
            to[out = 0, in = 225] (BSM.south west);
        \foreach \x in {0,...,2}{
            \pic at ($(BSMbw) - (0.5,0) - (0.3*\x,0)$) {Pulse=1};
            \pic at ($(BSMbe) + (0.5,0) + (0.3*\x,0)$) {Pulse=1};
        }
        \draw[light] (together) -- ($(BSMbe)+(0.5,0)$) to[out = 180, in = -45] (BSM.south east);
        \draw[light] (QBB.west)  to[out = 180, in = 0] (together);
        \draw[light] decorate [decoration={loop shape, loops/extra loops = 1, loops/side=-1}] 
            {(QBB.north west) -- ($(QBB.north west -| together)+ (0.25, 0)$) } 
            to[out = 180, in = 0] (together);
        \draw[light] decorate [decoration={loop shape}] 
            {(QBB.south west) -- ($(QBB.south west -| together)+ (0.25, 0)$) } 
            to[out = 180, in = 0] (together);
        \foreach \anchor in {north west, west, south west}{
            \pic at ($(QBB.north west)!(QBB.\anchor)!(QBB.south west) - (0.5,0)$) {Pulse = 1};
        }
    \end{tikzpicture}

    \vspace{1 ex}
    \begin{tikzpicture}
        \node[node font = \small] (label) at (-1/4,1.5) {$c)$};
        \node[server] (QBB) at (6,0) {};
        \foreach \i in {-1,0,1}{
            \draw ($(QBB) + (\i*0.2,0)$) pic [rotate = 195+\i*180, transform shape] {Spin};
        }
        \foreach \i in {-1,0,1}{
            \node[client] (cl\i) at (0,-\i*0.75) {};
            \pic[scale = 0.25, transform shape] at (cl\i) {laser};  
            \node[BSM, color = red!50!black] (BSM\i) at ($(cl\i.east)!0.8!(QBB.west|-cl\i.east) + (0,0.25)$) {};
        }
        \coordinate (split) at ($(cl0.east)!(BSM0.south east)!(QBB.west)!0.7!(QBB.west)$);
        \draw[color = red!50!black] (QBB.west)  to 
            pic {Pulse=3}
            (split);
        \foreach \i in {-1,0,1}{
            \draw[light] (split) to [out = 180, in = 0] ($(BSM\i.south east|-cl\i.east)+(0.5,0)$) to [out = 180, in= -45] (BSM\i.south east);
            \draw[light] (cl\i.east) --  ($(BSM\i.south west|-cl\i.east)-(0.5,0) $) foreach \j in {1, ..., 6}{
                pic [pos = \j/7] {Pulse=1}
            } to[out = 0, in = -135] (BSM\i.south west);
        }
    \end{tikzpicture}

    \vspace{2 ex}
    \begin{tikzpicture}
        \node[node font = \small] (label) at (-1/4,1) {$d)$};
        \node[client] (cl) at (0,0) {};
        \pic[scale = 0.25, transform shape] at (cl) {laser};
        \draw[light] (cl.east)  to 
            pic {Pulse=3}
            ++(0.5,0) coordinate (split);
        \foreach \i in {-1,0,1}{
            \node[server] (QBB\i) at (0.8*6-9/40,-2.4-0.92*\i) {};
            \foreach \j in {-1,0,1}{
                \draw ($(QBB\i) + (\j*0.2,0)$) pic [color= red!70!black, rotate = 195+\j*\i*180, transform shape] {Spin};
            }
            \node[BSM] (BSM\i) at (0.8*6-9/40,\i*0.75+0.25) {};
            \draw[light] (split) to[out = 0, in = 180] ++(0.4,0.75*\i) coordinate (help\i) -- ($(help\i-|BSM\i.south west)-(0.5,0)$) foreach \j in {1, ..., 6}{
                pic [pos = \j/7] {Pulse=1}
            } to [out = 0, in = -135] (BSM\i.south west);
            \draw [rounded corners, light] (QBB\i.east) [rounded corners] -- (6+1/8+1/4*\i,0 |- QBB\i.east) |- ($(help\i-|BSM\i.south east)+(0.75,0)$) pic [pos = 1,sharp corners] {Pulse=1} -- ++(-0.25,0) to [out = 180, in = -45] (BSM\i.south east);
        }
        \node[BSM,rotate = 90] (BSME0) at ($(QBB0.west)-(3,0)$) {};
        \coordinate (xBSME) at ($(QBB0.west)-(1.5,0)$);
        \foreach \i in {-1,1}{
            \coordinate (yBSME\i) at ($(QBB\i.west)!0.5!(QBB0.west)$);
            \node[BSM, rotate = 90] (BSME\i) at (xBSME |- yBSME\i) {};
        }
        \draw[light] (QBB-1.north west) -- ($(BSME0.south east|-QBB-1.north west)+(0.5,0)$) pic [pos = 1/2] {Pulse=1} to [out = 180, in = 45] (BSME0.south east) ;
        \draw[light] (QBB1.south west) -- ($(BSME0.south west|-QBB1.south west)+(0.5,0)$) pic [pos = 1/2] {Pulse=1} to [out = 180, in = -45] (BSME0.south west) ;
        \draw[light] (QBB0.south west) -- ($(BSME1.south east|-QBB0.south west)+(0.4,0)$) pic [pos = 1/2] {Pulse=1} to [out = 180, in = 45] (BSME1.south east) ;
        \draw[light] (QBB0.north west) -- ($(BSME-1.south east|-QBB0.north west)+(0.4,0)$) pic [pos = 6/8] {Pulse=1} to [out = 180, in = -45] (BSME-1.south west) ;
        \draw[light] (QBB-1.south west) -- ($(BSME-1.south east|-QBB-1.south west)+(0.4,0)$) pic [pos = 6/8] {Pulse=1} to [out = 180, in = 45] (BSME-1.south east) ;
        \draw[light] (QBB1.north west) -- ($(BSME1.south west|-QBB1.north west)+(0.4,0)$) pic [pos = 1/2] {Pulse=1} to [out = 180, in = -45] (BSME1.south west) ;
  
    \end{tikzpicture}
    \caption{Four schematic illustrations of multiplexing between two parties. a) Remote state preparation between a client equipped with a laser (left) and a node equipped with a quantum memory (right) is performed using an intermediate measurement station as in \cite{vanDam2025}. Traditional time-bin multiplexing allows many attempts to be performed provided one can fit more pulses in the beamline. b) Quantum multiplexing between two nodes (left and right) with asymmetric available memories enable higher entanglement generation rates. c) Quantum multiplexing between several clients (left) and one server (right) allows the server to provide higher rates of remote state preparation in the face of high demand. d) Multi-server multiplexing improves the rate at which a single client (left) can perform $s$-qubit remote state preparation on a collection of quantum server nodes (right) due to fast inter-server connections, especially for color centers where entanglement generation with a node degrades the quality of qubits already stored in that node.}
    \label{fourmults}
\end{figure}

A quantum internet will enable quantum secured applications for and between end users going beyond what can be done with a single quantum computer \cite{Wehner2018,Lee2022}, including including blind quantum computation (BQC) \cite{raussendorf2000quantum,briegel2009measurement,Broadbent2009,Fitzsimons2017}, private distributed quantum sensing \cite{Bugalho2025}, secure position verification \cite{Das2021}, and quantum key distribution \cite{Shor2000}. High qubit transmission rates are vital to many-qubit quantum internet experiments because quantum information is intrinsically fragile \cite{Terhal2012}, degrading in time due to decoherence \cite{Zurek1991}. Infidelic storage of qubits introduces a cost to storing qubits, either in terms of the fidelity of the reconstituted state or in a quantum error correction overhead \cite{Shor,Lidar2013,Terhal2015}. Commonly, one introduces cutoffs to quantum network protocols such that after a time window of $w$ attempts passes qubits are discarded, improving the fidelity at the cost of decreased protocol execution rate \cite{Li2020,Iesta2023,Davies2024}. End users possess either small quantum computers or photonic clients: devices with low quantum capabilities enabling transmission of many qubits to untrusted remote quantum servers, either directly or via quantum repeater chains that distribute entanglement over long distances \cite{Duan2001}.  Recently, various classes of photonic client designs have been proposed and explored both theoretically and experimentally \cite{Polacchi2023,Drmota2024,Iuliano2024}. While some of these client designs require the client to measure \cite{Drmota2024} or perform unitaries on single photons \cite{Polacchi2023}, others only require the client to emit a weak coherent pulse (WCP) \cite{Iuliano2024}. These same clients can implement the recently proposed single click and double-single click protocols for remote state preparation (RSP) \cite{vanDam2025}, which are analogous to the single click scheme for remote entanglement generation (EG) previously studied \cite{Campbell2008,Hermans2023,Beukers2024} and implemented  \cite{Kalb2017,Stolk2024}. 

These new single click protocols are also amenable to a qualitatively different form of multiplexing: quantum multiplexing, first introduced in Ref. \cite{LoPiparo2019} for many-qubit EG. Whereas classical multiplexing makes use of many signals to transmit information in parallel, quantum multiplexing makes use of coherence between spatial modes. This can make use of unutilized network resources in asymmetric connections and achieve an increase in the rate of RSP and EG. As we will show in section III, for example, by creating a superposition of zero and one photons and splitting it over many spatial modes \bea\label{spliteq}
 \xi\ket{\emptyset  } + \sqrt{1-\xi^2} \ket{1} 
\rightarrow \xi\ket{\emptyset  } + \sum_k\sqrt{\frac{1-\xi^2}{M}} \ket{1_k}
\eea one client or emitter in a repeater is able to interact with multiple quantum memories \emph{simultaneously}, improving the rate-fidelity tradeoff for both RSP and EG on top of what is achievable with classical multiplexing. This can implemented using simple linear optics (beam splitters, Bell state analyzers). For WCP clients, this is not a true superposition of zero and one photons, but also includes higher order terms degrading performance.

Trivially, WCP clients also possess the ability to interface with multiple quantum servers in a single shot, regardless of what RSP protocol they utilize; a coherent state $\ket{\alpha}$ can be transformed into a product of coherent states $\ket{\alpha/\sqrt{M}}^{\otimes M}$ using a beam splitter array. This enables parallelization of quantum experiments, and can greatly improve the rate of quantum protocol (i.e. BQC) execution for internetworked servers. 

Crucially, the two properties above (splitting a quantum state across modes and measurements, and interfacing a single device to multiple servers) enable multiplexing advantages for quantum protocols beyond what is usually possible in devices. Firstly, they allow asymmetric nodes (e.g. with inequal memory registers) to use quantum resources more efficiently. Secondly, they enable workarounds for situations where quantum information stored in a node is degraded by interaction with other qubits in that node, as is the case for color centers \cite{Reiserer2016,Ruf2021,deBone2024,Simmons2024}. 

This paper is organized as follows. We begin by introducing two semiclassical limits to multiplexing, using multiplexing with temporal modes on a WCP-based RSP protocol as an illustrative example. 
Next, we introduce single click quantum multiplexing and illustrate it through two  use-cases, both making use of single click protocols: EG between assymetric quantum nodes, and RSP between $M$ client devices (which may or may not be collaborating) and a single quantum node. Lastly, we introduce multi-server multiplexing and study RSP between one client and many servers as a use-case where WCP clients provide a novel benefit over other photonic client designs.\footnote{Additionally, WCP based clients enable asymmetric Bell state measurement placement as discussed elsewhere \cite{vanDam2025}. This gives a further advantage with the classical temporal multiplexing achievable; WCP-clients (being memoryless) don't need to wait for the Bell state measurement to finish before making another attempt. Since the server only needs to preserve the qubit long enough for confirmation of a success or failure from the Bell state measurement station, it can rapidly cycle the qubits used for failed RSP attempts increasing the number of temporal modes available for classical multiplexing.}

Throughout this paper, we will study the improvement introduced by multiplexing to the rate of execution of a quantum protocol. We quantify this by the multiplexing improvement factor $m_s$: the ratio of the multiplexed $s$-qubit protocol execution rate to the un-multiplexed $s$-qubit protocol execution rate. Mathematically, we define


\bea\label{improvementfactor}
m_s=\frac{R_{s|F}(M)}{R_{s|F}(1)}
\eea for a $s$-qubit protocol, with $R_{s|F}(M)$ the maximum rate achievable at a desired minimum final state fidelity $F$, $M$ the abstract multiplexing resource, and all functions implicitly depending on many experimental parameters. For single qubit protocols we drop the subscript and denote the advantage $m$. 

In this paper, we determine when $m_s$ can surpass the semiclassical limits introduced in the next section, and study the behavior of $m_s$ as $M$ is varied. In general, the linear case $m_s\propto M$ corresponds to fully parallelized independent experiments with identical resources without multiplexing, and serves as an upper-bound for the single qubit multiplexing improvement ($s=1$), but not for the many qubit equivalents ($s>1$) as we shall see shortly.

\section{semiclassical Multiplexing for Quantum Communications}

Classical multiplexing techniques for quantum signals are ubiquitous in quantum science and broadly applicable to quantum networks. There are extensive studies of the impact of and limits to multiplexing in e.g. parameter estimation \cite{Kay1993}, Bayesian experiment design \cite{Chaloner1995}, and machine learning \cite{Auer2002}, as well as in quantum repeater chains making use of various strategies \cite{Haldar2025}, quantum key distribution experiments \cite{Grunenfelder2021}, and quantum network/internet protocols in general making use of time-bin \cite{Dhara2021} and spectral \cite{Guha2015} multiplexing, especially in the presence of decoherence \cite{Collins2007}. However, to the authors knowledge there is no calculation of limits to the rate improvement (that is, $m_s$ as defined in Eq. \ref{improvementfactor}) achievable, especially in the presence of a finite window or minimum final state fidelity constraint for quantum communications protocols. 

Before introducing our quantum techniques and strategies, we begin by introducing classical limits that attempt answer the following question: Given an $M$-fold multiplexing of a physical signal, what is the maximum improvement to the quantum communication\footnote{With quantum communication protocols themselves of course forming subroutines of delegated quantum computation protocols.} protocol/application execution rate given a target fidelity achievable without classical or quantum correlations between attempts i.e. what is the maximum value of Eq. \ref{improvementfactor} without introducing additional quantum features?\footnote{E.g. quantum correlations between attempts, addressing-dependent decoherence rates.} We here first go through a concrete, illustrative example of temporal multiplexing and then introduce and discuss our semiclassical limits, leaving the full derivations to Appendix \ref{app:derive}.

\subsection{Example: Temporal Multiplexing for Remote State Preparation with Weak Coherent Pulses}

As an illustrative  warm up to quantum and multi-server multiplexing, we calculate the semiclassical limits Eq. \ref{Limit1Eq} and Eq. \ref{Limit2Eq} for a specific model: RSP between a client and server making use of temporal mode multiplexing. 

Remote state preparation (RSP) is a (theoretically) straightforward quantum protocol to securely prepare arbitrary quantum states on a server \cite{Bennett2001}. 
While in principle RSP can be used to create any quantum state, in practice preparing equatorial superposition states 
\bea\label{plustheta}
\ket{+_{\theta}} = \frac{1}{\sqrt{2}} \left(\ket{0} + e^{i\theta} \ket{1}\right)
\eea for $\theta\in\{\frac{k\pi}{4}\}_{k=0\dots7}$ is sufficient to enable BQC \cite{kapourniotis2023asymmetric}.\footnote{Notably, both direct transmission of a qubit and teleportation also enable BQC. However, these require the client to have one (or more) genuine qubits, wheras RSP does not require the client to have a qubit themself.} If an angle $\theta'$ is instead required by the client, the client transmits the angle $\theta'-\theta$ to the server, preserving the opacity of the qubit state. By preparing $s$ such states and performing CZ gates between them, a $s$-qubit graph state is formed. Furthermore, BQC can be made verifiably blind if the states can be prepared, stored, and manipulated with sufficiently high fidelity \cite{vbqc,leichtle2021verifying}, necessitating very fast RSP rates in the presence of significant decoherence.  

By using multiplexed temporal modes as in Fig. \ref{fourmults}a, it is possible to improve the rate of single qubit RSP and $s$-qubit RSP. We first consider batching $M$ RSP attempts. If a WCP client making use of the single click or double click RSP protocol, then temporal multiplexing does not require a memory overhead on the quantum node. 

 Without multiplexing, a single RSP attempt takes a time $\tau \sim 2L/c$, with $L$ the maximum distance from the client or server to the measurement station. Asymmetric measurement station placement next to the server enables $M$ attempts to be made within the time $\tau$ without requiring additional server memories, since the server knows whether the RSP attempt succeeded or failed before the client. Typically for WCP clients,\footnote{In principle one could try to overclock temporal multiplexing to such an extent that subsequent pulses overlap, but this will rapidly introduce errors degrading success probability and won't be considered here. Furthermore, in practice this is precluded by the fluctuation-dissipation theorem; qubits with small $T_1$ times may make for fast cycle times, but are bad for storing information unless other transitions are used.} one can fit at most $M \lesssim \tau\gamma$ independent pulses in the beamline, with  $\gamma=1/T_1$ the FWHM of the pulse(s). 

 If in the un-multiplexed case each attempt taking duration $\tau$ succeeds with probability $p$ in creating a state with minimum fidelity $ F$, then the rate achievable is $R_{|F}(1)=p/\tau$. Assuming that the multiplexed protocol takes the same duration $\tau$ to perform a batch of $M$ attempts (each of duration $\tau_M = \tau/M)$, that each attempt within the batch succeeds with probability $p$ and results in a state with the same minimum fidelity $ F$, and the batched attempt is successful if \emph{any non-zero number} of the $M$ batched attempts are successful, then the success probability of each batch is $1-(1-p)^M$ giving a maximum rate $R_{|F}(M) = [1-(1-p)^M]/\tau$. Thus we find 

 \bea\label{temporalmsingle}
 m = \frac{1-(1-p)^M}{p} \leq M
 \eea with equality with the upper bound only in the limit $p\rightarrow 0$. Physically, this upper limit $m=M$ corresponds to the case where there are $M$ independent copies of the same single-qubit protocol or experiment run in parallel, and demonstrates the power of multiplexing; we can achieve a speedup as if we had many copies of an experimental apparatus by reusing the protocol across temporal or spatial modes.
 

 We now consider the same protocol, but iterated to produce $s$ successful executions i.e. an $s$-qubit RSP state. Assuming noisy storage (pure dephasing characterized by a $T_2$ time), to achieve a given minimum final state fidelity $ F$ it is required that the $s$ successful attempts occur in a window $w$. Indeed, choosing a maximum window size $w$ is equivalent to picking a minimal accepted final state fidelity. For example, a simple worse-than-worst-case upper bound on $w$ is to consider \emph{all} $s$ RSP qubits decohere under a depolarizing channel for the full window time so that each qubit state undergoes a transformation $\rho\rightarrow e^{-w \tau /T_2} \rho + (1-e^{-w \tau /T_2})\frac{\idh}{2}$. Assuming each individual RSP results in a pure qubit state with fidelity $F_0$ before decoherence, this leads to a requirement $w\lesssim \frac{T_2}{\tau_e} {\rm Log} \left( \frac{(2F_0 -1)(1-2^{-s})}{F_{\rm min} - 2^{-s}}\right)$. As expected the window increases with the $T_2$ time of the memory, decreases with the duration of the attempt $\tau_e$, and goes to infinity as $F_{\rm min}\rightarrow\frac{1}{2^s}$.\footnote{To be clear, this trick of considering all $s$ qubits to undergo simultaneous decoherence for the full window $w$ is obviously unphysical, and is done purely to make the maths simple. In reality, only $s-1$ qubits are actually held in memory during RSP and at most only one can be generated in any given time step using temporal multiplexing.} 

 We can now reformulate our problem as a window problem: given a need to generate $s$ successful single-qubit RSPs within a window of $w$ attempts, how long must we wait for a successful $s$-qubit RSP to complete? This problem is well-studied in Ref. \cite{Davies2024}, and is not thought to be analytically solvable in generality. However, in the limit $p\rightarrow 0$, one must wait $\sim \frac{1}{{w-1 \choose s-1 } p^s}$ attempts on average.
 
 For the multiplexed case, approximately the same fidelity is achieved with the substitution $w\rightarrow Mw$ and $\tau\rightarrow \tau/M$; since we can make $M$ times as many attempts in the same laboratory clock time, we effectively increase our window by a factor of $M$ while rescaling the duration of each attempt.\footnote{This is actually an upper bound in performance, as we discuss in more detail in the appendix. We use this approach of putting the effect of multiplexing in the duration of each attempt and the size of the window---mathematically, de-multiplexing the arrivals---to make the problem amenable to analysis with tools from scan statistics \cite{Glaz2001,Davies2024}.} This is precisely the advantage that classical multiplexing enables: by making multiple attempts in a shorter time, we decrease the effective duration of each attempt. Using this approach of rescaling the duration and the window, we find one must wait $\sim \frac{1}{{Mw-1 \choose s-1 } p^s}$ attempts on average when multiplexing is used in the low-success probability limit $p\rightarrow 0$. With each individual attempt taking a time $\tau/M$ instead of $\tau$, we thus calculate our $s$-qubit multiplexing improvement factor in the $p\rightarrow 0$ limit

\bea\label{multiplexingimprovementexample}
 m_s &=& \frac{M {Mw-1 \choose s-1 }}{{w-1 \choose s-1 }}.
 \eea In the opposite limit $p\rightarrow 1$, we find $m_s = M$ independently of $s$. That $m_s$ is monotonic in $p$ is intuitive, but mathematically involved to show rigorously and left to further discussion in Appendix A. 
 
 Assuming a sufficiently stringent fidelity is chosen such that $w \gg s$, for $p\rightarrow 0$ we arrive at the tighter bound on the improvement factor $
 m_s \leq M^s$. In this limit, the overall process is Poissonian and we can give a nice physical interpretation; we would need $M^s$ independent (that is, not sharing qubits or resources) copies of an un-multiplexed experimental implementation of this RSP protocol to yield the same rate of $s$-qubit delivery that we can achieve with a single multiplexed implementation making use of $M$ temporal modes. This superlinearity illustrates the resource efficiency of semiclassical temporal multiplexing for many-qubit protocols.


\subsection{Semiclassical Limits to Multiplexing for Quantum Communications}

\begin{semiclassicallimit}\label{Limit1}
    For a single-qubit protocol or application over an effective channel, the multiplexing advantage is semiclassically bounded 
    \bea\label{Limit1Eq}
    m\leq M\leq M_c
    \eea with $M_c$ the number of end-to-end channels or copies of a single channel that can be used simultaneously in a given attempt, and $M$ the number that are utilized. 
\end{semiclassicallimit}  

This limit is straightforward to interpret; at best, using $M$ multiplexing resources (i.e. modes) yields an improvement to single-qubit protocol rates equivalent to having $M$ copies of the full experimental apparatus, and multiplexing resources are finite. In our previous example of temporal mode multiplexing from the prior section, $M_c \sim \tau \gamma$ is simply the number of pulses that can be fit into an un-multiplexed attempt time $\tau$. 

To derive the $s$-qubit generalization of Eq. \ref{Limit1Eq} we make use of techniques and results derived in Ref. \cite{Davies2024} for the study of the problem of finding the average number of attempts of a probabilistic process (e.g. RSP or EG) must be made on average to achieve $s$ successful executions of a protocol on a quantum network within a window of $w$ attempts, colloquially known as the ``window problem" of scan statistics \cite{Glaz2001}. We also assume the window size $w$ to be directly proportional to the coherence time $\tau_c$ so that generation of a state with a given fidelity is equivalent to generating a state within a specified window. This assumption of a single fixed window is crucial to the derivation of the second limit, and is the key assumption we exploit for multi-server multiplexing. Lastly, we assume that the rate achievable $R_{s|F}$ for a protocol improves more slowly with $p$ (the probability for any individual attempt to succeed) as the quality of the memory qubits increases, as detailed in Appendix \ref{app:derive}. 

\begin{semiclassicallimit}\label{Limit2}
    For an $s$-qubit protocol or application over an effective channel, the $s$-qubit multiplexing advantage is semiclassically bounded 
 \bea\label{Limit2Eq2}
m_s\leq M {Ms-1 \choose s-1} \leq M_c {M_cs-1 \choose s-1}
    \eea with $M_c$ the number of end-to-end channels or copies of a single channel that can be used simultaneously in a given attempt, and $M$ the number that are utilized. Furthermore, when the number of individual qubit attempts $w$ is much larger than the number of qubits $s$, $m_s$ is bounded by the more stringent limit 
    \bea\label{Limit2Eq}
    m_s\leq  M^s \leq M_c^s.
    \eea
\end{semiclassicallimit}  

Note that both Eq. \ref{Limit2Eq2} and Eq. \ref{Limit2Eq} are equivalent to Eq. \ref{Limit1Eq} for $s=1$. For all three expression (Eqs. \ref{Limit1Eq}-\ref{Limit2Eq}), we find the upper bound is only saturated when the success probability is asymptotically small, as we derive in Appendix \ref{app:derive} and illustrated in the prior section for temporal multiplexing. Importantly for the near-term, this means that the multiplexing improvement factor defined in Eq. \ref{improvementfactor} is \emph{maximal} for quantum networks with lower-quality rates and memories.

\section{Single Click Quantum Multiplexing}

We now turn our attention to the first beyond-classical multiplexing technique introduced in this paper: single click quantum multiplexing. Classical multiplexing relies on combining several classical or quantum signals in parallel or series. In contrast, we define quantum multiplexing as processes which utilize coherence between multiplexed quantum signals (photons, Fock number superpositions, or weak pulses) to improve performance. This definition is consistent with the original work on quantum multiplexing in Ref. \cite{LoPiparo2019}, where a single photon interacting with many memories can generate many-party entanglement via a reflection/transmission-based EG scheme \cite{Albrecht2013,Omlor2025}, which can then be distilled to achieve higher fidelity entanglement \cite{Deutsch1996}. This results in higher EG rates with fewer resources consumed than pair-wise EG \cite{LoPiparo2020,LoPiparo2020b}, and has been implemented experimentally \cite{Xie2023}. 

The double-single click remote state preparation protocol introduced in Ref. \cite{vanDam2025} is also a form of quantum multiplexing technique, albeit a more rudimentary one; here, coherence between multiple temporally-separated RSP attempts is exploited to remove a shared unknown phase and provide a speedup over the previously established double click RSP protocol. We can build upon this by exploiting coherence between parallel spatial modes for a further improvement. 

Notably, quantum multiplexing breaks the assumptions that are used in deriving the semiclassical limits from the prior section, paving the way for beyond-classical improvement as we will now demonstrate through two example use-cases: between two server/repeater nodes, and between many client devices and a single node in a quantum network. For both use-cases we consider quantum multiplexed versions of the single click protocols for EG \cite{Campbell2008,Polacchi2023} and RSP \cite{vanDam2025}. As we show in the coming section, these two protocols can be modified to improve performance between nodes with asymmetric memory availabilities as well as a single node providing service to multiple client devices.

\subsection{Use Case: Entanglement Generation Between Two Nodes in a Quantum Network}

\begin{figure*}
    \centering
    \includegraphics[width=0.9\linewidth]{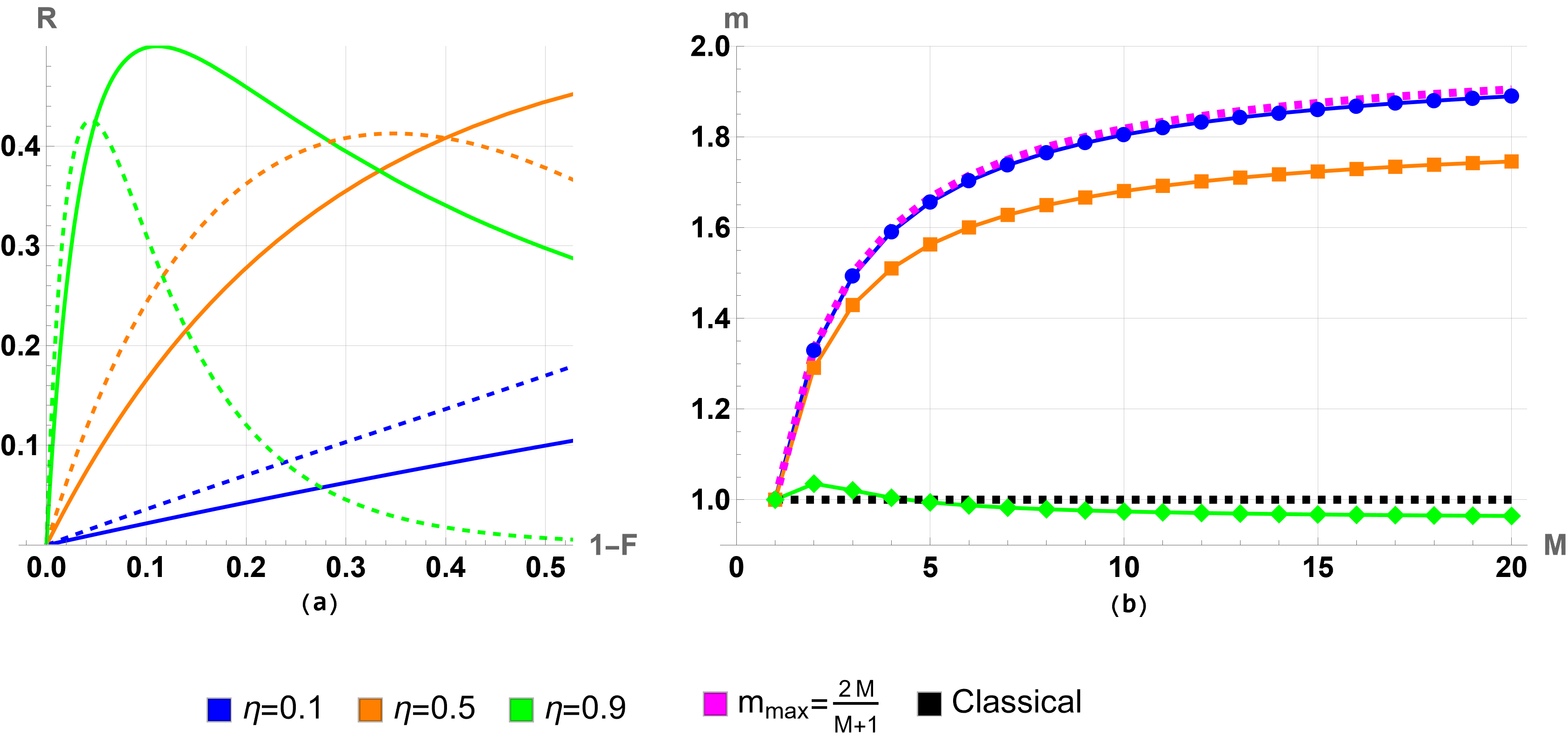}
    \caption{Single click quantum multiplexing between two nodes in a quantum network. (a) Fidelity-rate curves parameterized by $\xi_A$ for symmetric channel efficiencies $\eta_A=\eta_B=\eta$. Solid lines show $M=1$, dashed lines show $M=5$. The dimensionless rate given here is in units of the duration of an EG attempt. (b) Multiplexing gain $m$ as function of $M$ for generating link with minimum fidelity $F_\mathrm{min}=0.95$ between node $A$ and $B$ using $M$-to-$1$ quantum multiplexed heralded entanglement generation. The theoretical maximum in the small $\eta$ limit is $m_\mathrm{max}=\frac{2M}{M+1}$. Any $m\geq 1$ exceeds the classical bound. }
    \label{fig:use-case repeater}
\end{figure*}

We consider the situation in which two physically connected nodes in a quantum network wish to share an entangled link, but they have different numbers of quantum memories available for this task as in Fig. \ref{fourmults}b. We assume that node $A$ (right of Fig. \ref{fourmults}b) has $M>1$ available memories, while node $B$ (left of Fig. \ref{fourmults}b) only has $1$.\footnote{Note that here we do not differentiate between communication qubits and memory qubits as studied in Ref. \cite{Dam2017}. For such systems, the finite duration of SWAP gates creates a pragmatic limit to the applicability of single click quantum multiplexing, and our here-derived results only apply exactly in the fast-SWAP limit.} This situation could arise, for example, when the nodes of the network can dynamically reconfigure their memories to connect to different neighbours depending on the network's state (which could be realized through the use of an optical router as described in Ref. \cite{lee2022quantum}), or if one node is simply larger than the other. With a standard single click protocol the entanglement generation attempt can only use one out of the $M$ memories at $A$. The remaining $M-1$ would remain unused. Single click quantum multiplexing allows for all the memories at $A$ to be used and for a higher bright state/photon emission probability on $B$'s side. This gives rise to better performance for entanglement generation, as we will now show. 

 The $M$-to-$1$ single click quantum multiplexing entanglement generation attempt proceeds as follows. Each memory at node $A$, labeled $n=1,\dots,M$, emits a single photon when it is in the bright sate, which happens with probability $\xi_A$. The light is emitted into temporally separated modes, whose creation operators are denoted $a_n^\dagger$ satisfying canonical commutator relations $[a_n^\dagger,a_n]=1$, so that the light-matter state is 
 \bea\label{lossless}
 \ket{\psi}_A = \sum_{n=1}^M \sqrt{1-\xi_A^2}\ket{\zeros_n\emptyset}+\xi_A a^\dagger_n\ket{\ones_n\emptyset}, \eea with the first index corresponding to the $n$th qubit held in memory. 
 
 The modes are all coupled into the same optical fiber that transports them to a heralding station located in between nodes $A$ and $B$. The efficiency of the optical channel between node $A$ and the heralding station is given by $\eta_A\in [0,1]$. It includes all photon loss effects, such as coupling efficiencies and fiber attenuation. The state of the quantum memories and their corresponding optical modes at the time that they arrive at the heralding station is
\begin{multline}\label{eq: state that enters heralding from A}
    \rho_A =\\ \bigotimes_{n=1}^M\biggr[\left(\sqrt{1-\xi_A^2}\ket{\zeros_n\emptyset}+\sqrt{\eta_A}\xi_A a^\dagger_n\ket{\ones_n\emptyset}\right)(\mathrm{h.c.})
    \\+ (1-\eta_A)\xi_A^2 \ket{\ones_n\emptyset}\bra{\ones_n\emptyset}\biggr],
\end{multline}
where $\ket{\boldsymbol{0}_n}$ and $\ket{\boldsymbol{1}_n}$ are the qubit states of memory $n$ at $A$ and $\ket{\emptyset}$ denotes the photonic vacuum state.

At the same time that $A$ is emitting its pulses, the memory $B$ also emits a single pulse with bright state probability $\xi_B$ in a mode described by $b^\dagger_B$. When this pulse arrives at the heralding station, the combined state of the memory at $B$ and its optical mode is 
\begin{multline}\label{eq: state that enters heralding from B}
    \rho_B =\\ \left(\sqrt{1-\xi_B^2}\ket{\emptyset \zeros_B} + \sqrt{\eta_B}\xi_B b^\dagger_B\ket{\ones_B\emptyset }\right)(\mathrm{h.c.})\\ + (1-\eta_B)\xi_B^2 \ket{ \ones_B\emptyset}\bra{ \ones_B\emptyset},
\end{multline}
where $\eta_B$ is the efficiency of the optical channel between node $B$ and the heralding station. Now, the crucial step of the single click quantum multiplexing attempt is to split the photon from $B$ into an equal superposition of temporal modes that match the photons coming from the different memories of $A$. This step can be effected by using a splitter box that uses only linear optics, namely beamsplitters and delay-lines, to transform
\begin{equation}\label{eq: splitter box action}
    b^\dagger_B \to \frac{1}{\sqrt{M}}\sum_{n=1}^M b_n^\dagger,
\end{equation}
such that each mode described by $b_n^\dagger$ can interfere on a beamsplitter with its corresponding $a_n^\dagger$, as in the usual single click protocol. The interference is described by the beamsplitter transformation
\begin{align}\label{eq: beamsplitter transformation}
    a_n^\dagger &\to \frac{1}{\sqrt{2}}(c_n^\dagger + d_n^\dagger) \\
    b_n^\dagger &\to \frac{1}{\sqrt{2}}(c_n^\dagger - d_n^\dagger),
\end{align}
and the interference is followed by detections of output modes  with creation operators $c_n^\dagger$ and $d_n^\dagger$. The losses incurred in the splitter box are absorbed into $\eta_B$.

If out of all modes there is a single detection event at, say, $c_k^\dagger$, then this yields an entangled state between the memory $k$ at node $A$ and the memory at node $B$. More precisely, a single detection of mode $c_k^\dagger$ heralds the state
\begin{equation}\label{eq: state after click in ck abstract}
    \rho_{AB|c_k^\dagger}=\frac{\Tr_{n\neq k}[\bra{\emptyset}c_k\rho_{AB}c^\dagger_k\ket{\emptyset}]}{\Tr[\bra{\emptyset}c_k\rho_{AB}c^\dagger_k\ket{\emptyset}]},
\end{equation}
where the partial trace in the numerator is taken over all qubits at node $A$ different from $k$. 

In the lossless case, the numerator of \cref{eq: state after click in ck abstract} yields the unnormalized pure state
\begin{multline}\label{eq: lossless limit unnormalized}
    \ket{\psi_{AB|c_k^\dagger}} = \\\frac{\xi_B\sqrt{1-\xi_A^2}}{\sqrt{M}}\ket{\zeros_k \ones_B} + \xi_A \sqrt{1-\xi_B^2}\ket{\ones_k \zeros_B}.
\end{multline}
Through an appropriate choice of $\xi_A$ and $\xi_B$ this becomes the Bell state $\ket{\Psi_{AB}^+}=\frac{1}{\sqrt{2}}\left(\ket{\zeros_k \ones_B}+ \ket{\ones_k \zeros_B}\right)$. In the case with channel losses, the fidelity to $\ket{\Psi_{AB}^+}$ is given by
\begin{align*}
    F &= \bra{\Psi_{AB}^+} \rho_{AB|c_k^\dagger}\ket{\Psi_{AB}^+} \\
    &=\frac{1}{2}\frac{\left(\xi_B\sqrt{\eta_B(1-\xi_A^2)} + \xi_A\sqrt{M\eta_A(1-\xi_B^2)}\right)^2}{ \xi_B^2\eta_B(1-\eta_A\xi_A^2)+\xi_A^2M\eta_A(1-\eta_B\xi_B^2)}.
\end{align*}
Given a value for $\xi_A$, the fidelity is maximized by setting 
\begin{equation}
    \xi_B^2 = \frac{\eta_A\xi_A^2 }{\eta_B/M + (\eta_A-\eta_B/M)\xi_A^2}.
\end{equation}
Finally, the denominator in \cref{eq: state after click in ck abstract} is the probability that the detector clicks in the mode described by $c_k^\dagger$. The probability to obtain exactly one click in any of the $2M$ output modes is then
\begin{multline}\label{eq: probability single click}
    \Pr[\mathrm{single\,click}] = 2M \Tr[\bra{\emptyset}c_k\rho_{AB|c^\dagger_k}\ket{\emptyset}]
    \\ 
    = M\eta_A\xi_A^2(1-\eta_A \xi_A^2)^{M-1}(1-\eta_B\xi_B^2) \\+ \eta_B\xi_B^2(1-\eta_A\xi_A^2)^M.
\end{multline}
The rate at which an entangled state is generated in the $M$-to-$1$ single click quantum multiplexing protocol is 
\begin{equation}\label{eq: rate M to 1 q mxing}
    R = \frac{\Pr[\mathrm{single\,click}]}{\tau_e},
\end{equation}
where $\tau_e$ is the time needed for a single attempt. Note that in remainder of the analysis we will set $\tau_e=1$ so that the rate is dimensionless. The rate and maximized fidelity are shown as curves parameterized by the brightstate parameter $\xi_A$ in \cref{fig:use-case repeater}a. 

The multiplexing gain $m$ for generating an entanglement link with minimum fidelity $F_\mathrm{min}$ between nodes $A$ and $B$ using $M$-to-$1$ single click quantum multiplexing as described above is shown in \cref{fig:use-case repeater}b for symmetric channel efficiencies ($\eta_A=\eta_B\equiv \eta$).  For low channel efficiencies $\eta$, single click quantum multiplexing improves the rate as the number of memories increase, and closely follows the maximum value for high loss ($\eta=0.1$). We also see that for high channel efficiency ($\eta=0.9$) the rate is decreased by single click quantum multiplexing except for very high fidelity target states, so that single click quantum multiplexing is not generally beneficial. This is because single click quantum multiplexing requires a larger bright state probability to be used, and this quantity cannot exceed unity.

We find that in the high-loss regime $m\to 2$ as $M\to \infty$. Indeed, in this regime the fidelity is maximized for small bright state bright state probability so that we can expand $R$ and $F$ in $\xi_A^2$. Lowest order expansions yield
\begin{equation}
    R = 2M \eta_A \xi_A^2 
\end{equation}
and
\begin{equation}
    F = 1-\frac{M\eta_A(1-\eta_B)+(1-\eta_A)\eta_B}{2\eta_B}\xi_A^2.
\end{equation}
In this regime, the rate can thus be expressed as a function of the fidelity as
\begin{equation}\label{eq: rate as function of fidelity repeater small eta}
    R(M,F) = 4(1-F)\left[\frac{1-\eta_B}{\eta_B}+\frac{1-\eta_A}{M\eta_A}\right]^{-1}.
\end{equation}
In the symmetric case, $\eta_A=\eta_B=\eta$, and in the limit of small channel efficiency $\eta$ (that is, high loss), the $M$ dependence of $m$ is then given by 
\begin{equation}\label{eq: symmetric quantum multiplexing gain}
    m(M) \sim \frac{2M}{M+1}.
\end{equation}
Single click quantum multiplexing can thus improve over classical multiplexing by at most a factor of $2$ in the symmetric case. However, adding even a single additional qubit ($M=2$) provides a $50\%$ improvement to the rate of EG achievable in the high-loss regime. 

On the other hand, when the channel efficiencies are not the same, then for small efficiencies $\eta_A$ and $\eta_B$ we find from \cref{eq: rate as function of fidelity repeater small eta} that 
\begin{equation}\label{limitofmultiplexingq}
\lim_{M\to \infty} m(M) = \frac{\lim_{M\to \infty}R(M,F)}{R(1,F)} \sim 1+\frac{\eta_B}{\eta_A}.     
\end{equation} 
We observe that the single click quantum multiplexing improvement can be larger than $2$ only if the link on the side with only one memory (B) is more efficient than the link on the side with many memories (A).\footnote{Notably, in the opposite limit where the measurement station is next to the node with many memories such that $\eta_B\ll\eta_A$, semiclassical multiplexing improves the rate by a factor $M$ simply by resetting the memories after each unsuccessful attempt. Indeed, in the extremity of this limit where $\eta_A\rightarrow 1$ single click quantum multiplexing yields no improvement whatsoever as we can see from Eq. \ref{limitofmultiplexingq}.} In particular, we see that the efficiency of the connection to node $B$, which holds one memory, remains a bottleneck in the infinite $M$-limit; as $M$ grows large the probability to receive a photon from node $A$ also becomes large, but the probability to receive a photon from $B$ must remain the same. We see that single click quantum multiplexing is especially useful when it is used to mitigate a low efficiency on the side with many qubits (A), and indeed diverges as $\eta_A \rightarrow 0$ due to the un-multiplexed rate going to zero quickly. 

This example makes clear that single click quantum multiplexing is especially useful when the connection to the few-memory node is efficient, and the connection to the many-memory node is inefficient. By increasing the number of memories on the inefficient node we can then boost performance. This could possibly be exploited in a network resource allocation policy; the nodes closest to the heralding station should use little memories for entanglement generation, while the nodes far away from the heralding stations should use many. For example, if one has $10$ memories to distribute between two nodes, we can now see that the optimal distribution depends on the ratio of losses, varying between $5$ and $5$ (symmetric loss) and $1$ and $9$ (asymptotically asymmetric loss). We emphasize that, given a constant total loss $\eta_A\eta_B$ (corresponding to a constant total distance between nodes), an optimal rate is always achieved through a symmetric distribution of loss (symmetric Bell state analyzer placement). The single click quantum multiplexing strategy is best suited for cases where either asymmetric loss is unavoidable e.g. quantum networks with different physical platforms with different transduction requirements, asymmetric distances to the Bell state analyzer, or in networks where there are already (unutilized) asymmetric memories. 

Our result of an asymptotic constant multiplicative speedup in \cref{eq: symmetric quantum multiplexing gain} is consistent with the results found in Ref. \cite{Collins2007}. Here, symmetric emitter-by-emitter multiplexing can have a linear speedup within a single repeater link. This provides an upper bound on the speedup as it corresponds to having $M$ identical copies of the system.\footnote{At least, without utilizing more advanced multi-mode generalized Bell state measurements as in Ref.~\cite{Niv2024} or tree-state qubit encodings of qubits as in Ref.~\cite{Borregaard2020}.} Notably, Ref.~\cite{Collins2007} also shows that the speedup to the total $N$-repeater chain is polynomial in $M$, with the degree of the polynomial given $2^N - 1$, pointing towards repeater-chain generalizations of our result above.

In the near term, quantum multiplexing experiments are feasible using nodes with multiple emitters and phase stabilized optical fiber connections, and have already been implemented in Ref. \cite{Xie2023} for the reflection based quantum multiplexing EG scheme proposed in Ref. \cite{LoPiparo2019}. For example, the multiplexed entanglement generation implemented experimentally in Ref. \cite{Ruskuc2025} could be modified using single click quantum multiplexed entanglement generation to improve entanglement generation further for the asymmetric case.\footnote{Notably, in \cite{Ruskuc2025} they consider preparing multi-qubit $W$ states via multiple entanglement generation attempts. We can imagine further generalizing our single click quantum multiplexing method to prepare three-qubit $W$ states in a single attempt by postselecting on two clicks. Such an implementation of this would be a stepping stone to generalized Bell state measurements with tree states  \cite{Niv2024} and teleportation of logical qubits using more advanced parallelized encodings \cite{Borregaard2020}.}

\subsection{Use Case: $M$-Client Remote State Preparation with a Single Server Node}

\begin{figure*}
     \centering
     \includegraphics[width = \textwidth]{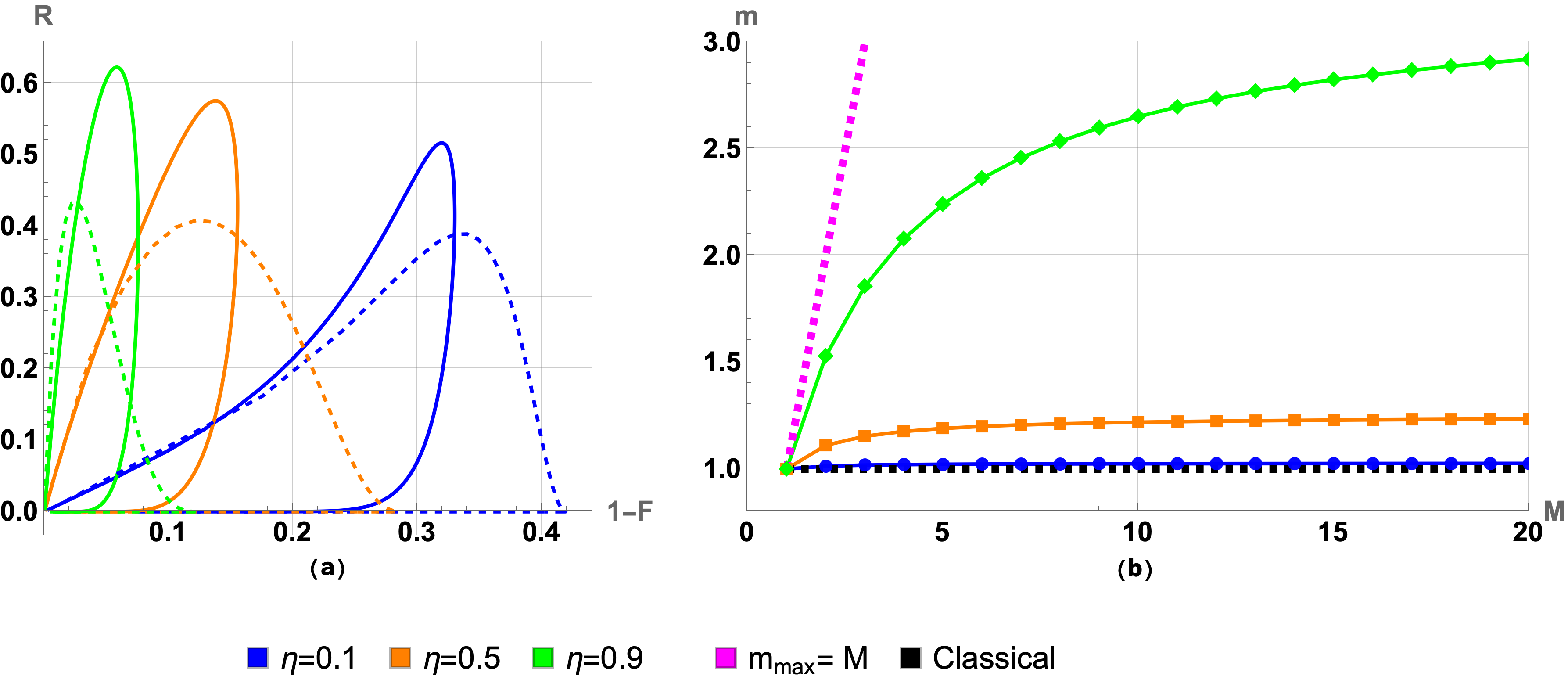}
    \caption{Single click quantum multiplexing between several client devices held by a single client and a single node in a quantum network with a continuous demand for single-qubit RSP. (a) Fidelity-rate curves parameterized by $\gamma=\eta_c |\alpha|^2$. Solid lines show $M=1$, dashed lines show $M=5$.  The dimensionless rate given here is in units of the duration of an RSP attempt. (b) Multiplexing gain $m$ as function of $M$ for RSP with fidelity $F_\mathrm{min}=1-10^{-6}$ using $M$ clients. The theoretical maximum with a perfect server and the classical bound is plotted using the magenta and black dashed lines, respectively. Similar results are observed for modified network protocols where the client devices are held by different users with continuous demand (Fig. \ref{fig: fidelity rate curves for clients 2}) as well as one-time demand (Fig. \ref{fig: single use}) are calculated and discussed in Appendix B.}
    \label{fig:use-case client}
\end{figure*}

We now turn our attention to single click quantum multiplexing between $M$ photonic client devices and a single node attempting single click RSP \cite{vanDam2025}, as despcted in Fig. \ref{fourmults}c. For simplicity here we consider a situation where the $M$ photonic client devices are used simultaneously by a single user. In Appendix \ref{app:RSP appendix} we consider the case where each of the $M$ client devices correspond to different users competing for access to the node, with near-identical results.  

The client devices (left of Fig. \ref{fourmults}c) each emit a WCP with the same mean photon number $|\alpha|^2$ but choose separate phases. After introducing channel loss $1-\eta_c$, the state of the clients are
\begin{align}\label{emil1}
    \ket{\psi_c} = \bigotimes_{m=1}^M\ket{\sqrt{\eta_c}\abs{\alpha}e^{-i\theta_m}}.
\end{align}
Variances in channel losses among the clients can be compensated by changing the mean photon number such that $\eta_c|\alpha|^2$ is the same for all client devices.

The server (right of Fig. \ref{fourmults}c) will split their signal into the $M$ channels corresponding to the $M$ client devices, each with (identical) transmission efficiency $\eta_s$. This results in a state
\begin{align}
    \rho_S = \left(\sqrt{1-\xi^2}\ket{0\emptyset}+\sqrt{\frac{\eta_s}{M}}\xi\sum_m\ket{11_m}\right)\left(\text{h.c.}\right) \nonumber\\+ (1-\eta_s)\xi^2\ket{1\emptyset}\bra{1\emptyset},
\end{align}
where $\xi$ is the bright state probability of the server qubit. We will now condition on a successful detection in \emph{one and only one} of the $M$ channels without photon number resolution. Thus, we get the transformation 
\begin{align} \label{eq:dm_rsp}
    \rho_{RSP|k+} = \frac{\sum_{n=1}\Tr_{m}[\bra{n_{k+}}\rho \ket{n_{k+}}]}{\sum_{n'=1}\Tr[\bra{n'_{k+}}\rho \ket{n'_{k+}}]},
\end{align}
where $\Tr_{m}$ is over the photonic modes, $\rho=\ket{\psi_c}\bra{\psi_c}\otimes\rho_S$ and $\ket{n_{k+}}$ is the state with $n$ photons in the ``plus" detector mode of the $k$'th channel. We leave the details to the remaining calculation in appendix \ref{app:RSP appendix}. After maximizing the fidelity over the bright state probability we find 
\begin{widetext}
\begin{align} \label{eq:Fid_Rsp}
    F&= \frac{1}{2}\left(1+\sqrt{\frac{\eta_s\eta_c|\alpha|^2/M}{\pqty{1-e^{-\eta_c|\alpha|^2/2}}\bqty{1-e^{-\eta_c|\alpha|^2/2}+\eta_s\pqty{\frac{2+\eta_c|\alpha|^2}{4M}+1-e^{-\eta_c|\alpha|^2/2}}}}}\right) \\\nonumber
    &\sim 1-\frac{M(1-\eta_s)+\frac{\eta_s}{4}}{4\eta_s}\eta_c|\alpha|^2,
\end{align}
where the last line is the small-$\alpha$ limit. 
As in the previous analysis, we define the duration $\tau_e$ of each attempt $\tau_e$ to be unity so that our rate calculated is the same as the probability to get a single click in any of the detectors
\begin{align} \label{eq:Rate_Rsp}
    R= \frac{4e^{-(M-\frac{1}{2})\eta_c|\alpha|^2}\pqty{1-e^{-\eta_c|\alpha|^2/2}}\bqty{4M\pqty{1-\eta_s}\pqty{1-e^{-\eta_c|\alpha|^2/2}}+\eta_s\pqty{2+\eta_c|\alpha|^2}}}{4(2-\eta_s)\pqty{1-e^{-\eta_c|\alpha|^2/2}}+\frac{\eta_s}{M}\pqty{2+\eta_c|\alpha|^2}} \sim 2M\eta_c|\alpha|^2.
\end{align}
\end{widetext}
where the last form given is the small-$\alpha$ limit. Note that in practice, the small-$\alpha$ limit is the regime of interest since the chance that a malicious server determines the state of the RSP qubit increases with the mean photon number $|\alpha|^2$ for RSP protocols with WCP clients \cite{Garnier2024}.

We have plotted the rate for our single user controlling all $M$ devices against the fidelity obtained in Fig. \ref{fig:use-case client}a. This leads to a mulitplexing advantage, but only in the high-fidelity limit:

\bea\label{mfactorclient1}
m(M) &\sim& \frac{M (1+\frac{4}{\eta_s} -4)}{1+\frac{4M}{\eta_s} -4M} \nonumber \\ 
&\rightarrow& 1+\frac{\eta_s}{4(1-\eta_s)}
\eea where in the second line we take the $M\rightarrow \infty$ limit. We plot the multiplexing factor in Fig. \ref{fig:use-case client}b, and observe it indeed increases with $\eta_s$. We get the maximum achievable multiplexing gain in the limit as $\eta_s\to 1$, and in the high fidelity regime, which yields $m(M)\to M$. 

Intuitively, it is surprising that collaborating client devices yields any advantage in any regime, as collaborating client devices are equivalent to a single client device and a beam splitter, up to phase modulations. However, the additional spatial modes reduce the likelihood of an undetected multi-photon emission; as the probability of such an error goes as $\sim 1/M$. This gives a slight improvement of the fidelity at the same value of $\eta_c|\alpha|^2$. When increasing $\eta_c|\alpha|^2$, the probability of detections in multiple channels increase, which are marked as unsuccessful heralding patterns. Thus, after a critical value the rate will decrease when increasing $\eta_c|\alpha|^2$ if the measurement stations do not have photon number resolving detectors, as seen in Fig. \ref{fig:use-case client}a (but not in Fig. \ref{fig:use-case repeater}a, where there were no multi-photon events). However, in the high fidelity regime before the critical value, $\eta_c|\alpha|^2\ll 1$ and the improvement in fidelity has a greater impact than the decrease in rate due to multi-client photons emitted.



We emphasize that here we have omitted a detailed discussion of security for WCP based RSP, which places further constraints on the values of $|\alpha|^2$ that can be used in any practical implementation \cite{Garnier2024}. However, something interesting has happened in this protocol; as we only have one resource available at the server side, here multiplexing in the semiclassical method is impossible. By utilizing the quantum properties of the single photon from the quantum side, we can improve performance far beyond the semiclassical multiplexing limit.  Furthermore, we note that there exists no upper limit to the multiplexing gain for a lossless server. While this is not remotely realistic in the near-future, it illustrates the potential benefits of a quantum approach to multiplexing.


\section{Multi-Server Multiplexing}

In the previous section, we introduced single click quantum multiplexing as a new method for improving the rate of qubit transmission through a network by a factor $m$, illustrated through two use-cases. In this section, we now introduce a method for multiplexing where the achievable speedup $m_s$ to the $s$-qubit transmission through a network is much greater, surpassing the semiclassical limit Eq. \ref{Limit2Eq2}. The reason for this is twofold; firstly, the connections within a quantum network are not in general homogeneous in terms of quality. Secondly, interactions (e.g. entanglement generation, remote state preparation) with a server may degrade the quality of qubits stored therein (e.g. as is the case for color centers \cite{Chakraborty2022}); in the language of the window problem of scan statistics introduced in section II, this means that utilizing many servers to perform a task that, strictly speaking, only requires a single server, may increase the effective window, i.e., number of attempts that can be made without discarding qubits held in storage, as was first emphasized in Ref. \cite{Collins2007}.

Unsurprisingly, this large improvement to the $s$-qubit rate comes at a cost; multi-server multiplexing between a single WCP photonic client and $M$ interconnected nodes in a quantum network is the most resource-intensive use-case we study. We now study this situation in detail, and utilize a simple toy model to both analytically derive $m_s$ for $s=2$ and numerically calculate $m_s$ for $s>2$ for our proposed use case of $s$-qubit RSP, which is a prerequisite for BQC.

\subsection{Use Case: $s$-Qubit Single-Client RSP} 

As a final use-case, we consider multi-server multiplexing between a single WCP photonic client attempting $s$-qubit RSP with a hub of $M$ server nodes as illustrated in Fig. \ref{fourmults}d. Here, we can use multi-server multiplexing to attain beyond-classical scaling in  regimes with fast/efficient inter-server links and memories that are more susceptible to decoherence during EG/RSP attempts (i.e. color centers). 

Unlike the multiplexing based on modifying the single click entanglement generation \cite{Hermans2023,Stolk2024} or remote state preparation \cite{vanDam2025}, multiplexing a WCP photonic client to multiple nodes for remote state preparation doesn't require us to do anything particularly quantum; we simply put our pulse on a beam splitter array so that pulses are transmitted to $M$ nodes in parallel, and increase the power of the laser by a factor of $M$ to compensate.\footnote{This of course will change the chance of leaking the value of $\theta$, but this is straightforward to fix by randomly rotating the phases from each port, generating a one-time pad. Note that for information theoretic security of RSP with WCPs a physical implementation of the ``security gadget'' proposed in Ref.  \cite{Garnier2024} must be utilized. In practice, this is accomplished by preparing additional qubits and performing CZ gates with the output of the $s$-qubit RSP, forming a one-time pad via a measurement-based rotation.} For single click, double click, and double-single click (in batches of $K$ pulses), this requires a substitution 

\begin{widetext}
\bea\label{MNodesSCDC}
\ket{\alpha e^{i\theta}} \rightarrow \ket{\alpha e^{i\theta_k}}^{\otimes M}\,\,&{\rm (Single \, Click)} \nonumber  \\
\ket{\alpha/\sqrt{2}, \alpha/\sqrt{2}e^{i\theta}} \rightarrow \ket{\alpha/\sqrt{2}, \alpha/\sqrt{2} e^{i\theta_k}}^{\otimes M} \,\,&{\rm (Double \, Click)}\nonumber \\
\ket{\alpha e^{i\theta_k}}^{\otimes K} \rightarrow \ket{\alpha e^{i\theta_k}}^{KM}  \,\,&{\rm (Double{\rm -}Single \, Click)}\nonumber \\
\eea \end{widetext} where implicitly the phase $\theta_k$ is assumed to be modulated to different values over spatial (and for double-single click, temporal) modes. 

Increasing the amplitude within any particular mode introduces errors due to multi-photon events and degrades fidelity. However, here the additional multi-photon events are detected in different spatial modes and can be rejected without photon number resolution, as was the case in section IIIB.

Since the fidelity is unchanged by multiplexing, it is trivial to calculate the single-qubit improvement factor $m$ for RSP to \emph{any} node in a symmetric network configuration: it is simply given by

\bea\label{mqbpsMserverSimp}
m = M (1-p)^{M-1} \leq M. 
\eea Here, $p$ is the success probability of a single RSP attempt (made using either double click or single click protocol), and $M$ is the number of nodes. Notably, this expression also describes attempting RSP between a single client and $M$ emitters within a single node, and the upper bound is saturated in the low-$p$ limit. Indeed, this is just a more resource-intensive implementation of the semiclassical limit-saturating speedup seen for temporal multiplexing in section IIA.

In this section, our main subject of interest is the speedup achieved in preparing $s$-qubits on a single arbitrary node in a quantum network. Now we must contend with decoherence during the storage of the qubit(s) as well as the interconnectivity of the network itself. This is, in general, a non-trivial problem to solve. Recent work on color centers \cite{Reiserer2016,Ruf2021,deBone2024,Simmons2024} shows that addressing a color center tends to dramatically reduce the qubit lifetimes of other qubits stored in the same color center (which typically contain multiple memory qubits per communication qubit). This gives rise naturally to several simple multiplexing strategies to consider, which we denote $\sigma$. The simple strategies we consider are ``try and commit" ($\sigma = t$) where a client interacting with $M$ nodes commits to the first node that yields a success, and ``multiplex" ($\sigma = m$) where while there is a qubit stored in a node no attempt is made by the client to remotely prepare a qubit on it. This list is by no means exclusive, but suffices to illustrate where the full advantage of multi-server multiplexing lays.\footnote{For example, we could also consider a fully multiplexed strategy where at every instance the client attempts to remotely prepare a state on every node, including those that already hold qubits. Since this combines the downsides of both approaches for our toy model we omit it in this analysis.}

We begin by tackling the case of $s=2$ analytically before moving on to $s>2$ numerically. In both cases we find beyond-classical improvement $m_s \geq m\times M^s$. 

\subsection{Toy Model for $s=2$: Analytic Solutions}

\begin{table*}
    \centering
    \begin{tabular}{|c|c|c|}
    \hline 
        \textbf{Protocol} & \textbf{Success Probability} & \textbf{Fidelity}\\
        \hline  \hline  
        Single Click RSP & $2\eta_c|\alpha|^2$ & $ 1-\frac{\eta_c(4-3\eta_s)}{16\eta_s}|\alpha|^2$\\
          Single Click EG & $2\eta_A \eta_B \xi^2$ & $1-\xi^2$ \\
         Double Click RSP & $\frac{ \eta_c\eta_s}{2} |\alpha|^2$ & $1-\frac{\eta_c}{\eta_s}\frac{4-3\eta_s}{16}|\alpha|^2$\\
        Double-Single Click RSP & $\frac{4}{3}\eta_c|\alpha|^2$ & $ 1-\frac{\eta_c}{\eta_s}\frac{4-3\eta_s}{8}|\alpha|^2$ \\
        Measurement-Only RSP & $ \eta_s \eta_c $& $\approx 1$ 
        \\
        \hline
    \end{tabular}
    \caption{Summary of the optimized success probabilities $P$ and fidelities $F$ for the single click and double click remote state preparation (RSP), as well as single click entanglement generation (EG), protocols in the low-WCP amplitude $\alpha$, low bright state probability/photon emission probability $\xi$, and high loss limits \cite{vanDam2025}, as well as for the measurement-only remote state preparation protocol studied in \cite{vanDam2024}. Here we only consider loss in fidelity inherent to the protocol and not from other sources e.g. depolarizing in fiber, infidelitous server photons. We define $\eta_{A/B/s/c}$ as the efficiency for the left/right/server/client nodes, including the effects of loss due to transmission in fiber, and do not include any multiplexing (which improves the rate by a factor $m$ and leaves the fidelity unchanged, by definition). For a more in-depth review of single click entanglement generation see appendix \ref{app:SCreview}.}\label{tableprobs}
\end{table*}

We now introduce a simple toy model for $s=2$ qubit RSP on $M$ server nodes (Fig. \ref{fourmults}d). We consider a network configuration where each server-server/server-client link has an associated single-qubit RSP or entanglement generation rate $R_{ij}=\frac{P_{ij}}{\tau_e}$. This rate implicitly absorbs all single-qubit multiplexing effects as well as imperfections in the quantum optical implementation, and also explicitly assumes uniform attempt duration $\tau_e$ between RSP and EG. Here, $P_{ij}$ depends on the RSP or entanglement generation protocol used, which are summarized in the low bright state probability $\xi$ and low coherent state amplitude $\alpha$-limit in Table. \ref{tableprobs}.


We also associate with each nodes two coherence times $\tau_{ce}$ and $\tau_{co}$ corresponding to coherence time with ($e$) and without ($o$) EG/RSP attempts. These, along with other experimental factors (e.g. Nitrogen vacancy center qubit loss via charge state conversion necessitating charge-resonance checks \cite{Chakraborty2022}), impose two seperate cutoffs (i.e. finite attempt window sizes) $n_e$ and $n_o$ for each server node, again corresponding to with ($e$) and without ($o$) a qubit held at the server. Considering these two cutoff times separate instead of using a unified window $w$ is an approximation made for ease of calculation, but is reasonable for $n_o \gg n_e$ and $\tau_{co} \gg \tau_{ce}$. For our analytic toy model, we further consider a highly symmetric network such that there are only two success probabilities for attempts: server-server $P_{ss}$ and server-client $P_{sc}$ and assume uniformity otherwise. These yield analytic rates for the try and commit $\sigma=t$ and multiplex $\sigma = m$ strategies:

\begin{widetext}
\bea\label{Rates2M:t}
R^{2,M}_{\sigma=t} &=& \frac{1- (1-P_{sc})^{n_e}}{\frac{\tau_e}{M P_{sc}} +\frac{1-P_{sc}}{P_{sc}}\left(1- (1-P_{sc})^{n_e} - n_e P_{sc} (1-P_{sc})^{n_e}\right)\tau_e + (1-P_{sc})^{n_e}n_e\tau_e}  \\
R^{2,M}_{\sigma=m} &=&  \left(1- (1-P_{sc})^{(M-1)n_o}\right)\left(1- (1-P_{ss})^{n_e}\right) \nonumber \\
&/& \small \left( \frac{\tau_e}{MP_{sc}} + 
    \frac{1-P_{sc}}{P_{sc}}\left(1- (1-P_{sc})^{(M-1)n_o} - (M-1)n_o P_{sc} (1-P_{sc})^{(M-1)n_o}\right)\frac{\tau_e}{M-1} + (1-P_{sc})^{(M-1)n_o}n_o\tau_e \right. \nonumber \\
    &+& \left. (1-(1-P_{sc})^{(M-1)n_o})\left(\frac{1-P_{ss}}{P_{ss}}\left(1- (1-P_{ss})^{n_e} - n_e P_{ss} (1-P_{ss})^{n_e}\right)\tau_e + (1-P_{ss})^{n_e}n_e\tau_e\right) \right).\label{Rates2M:m}
\eea\end{widetext}

These expressions must be optimized over the parameters in the success probabilities $P_{ss/sc}$ and the cutoffs $n_{e}$ and $n_o$ (again, subject to physical constraints due to qubit loss) to maximize the two-qubit RSP rate subject to a minimum fidelity constraint. 

The infidelity has in general two components: an intrinsic infidelity from the RSP/EG protocol used as summarized in Table \ref{tableprobs}, as well as an infidelity due to decoherence during storage. For an $s$-qubit protocol

\bea\label{fidelitycalc}
F_{1,2,\dots s} = \frac{1}{2^s} \prod_i \left(1+\expect{e^{\sum_j-\frac{t_{ije}}{\tau_{ce}} -\frac{t_{ijo}}{\tau_{co}}}}_{t_{ije},t_{ijo}\forall j}\right)
\eea where $t_{ije/o}$ are the times a qubit is stored in memory of node $j$ with ($e$) and without ($o$) EG/RSP attempts made at the $j$th node. In the absence of an adaptive protocol we can assume separability of the expectation value 

\bea\label{expectationfidelity1}
\expect{e^{-\frac{t_{ije}}{\tau_{ce}} -\frac{t_{ijo}}{\tau_{co}}}}_{t_{ije},t_{ijo} \forall j} &=& \prod_j \expect{e^{- \frac{t_{ije}}{\tau_{ce}}}}_{t_{ije}}\expect{e^{-\frac{t_{ijo}}{\tau_{co}}}}_{t_{ijo}} \nonumber \\
\eea so that each expectation value is then a product of geometric series depending on $n_{e}$ or $n_{o}$ and the success probabilities conditioned on there being a single success, as failed attempts of the full protocol do not contribute to infidelity. Removing subscripts for readability and generality, we have in general,

\bea\label{expectationfidelity2}
\expect{e^{-\frac{t_{e/o}}{\tau_{ce/o}}}}_t &=& \sum_{n=1}^{n_{e/o}} P_{n|1}^{e/o} e^{-\frac{n\tau_e}{\tau_{ce/o}}} 
\eea with

\bea\label{expectationfidelity2}
 P_{n|1}^{e/o} &=& \frac{(1-P_{sc/ss})^{n-1} P_{sc/ss}}{\sum_{n=1}^{n_{e/o}}(1-P_{sc/ss})^{n-1} P_{sc/ss} } \nonumber \\
&=& (1-P_{sc/ss})^{n-1} \frac{P_{sc/ss}}{1-(1-P_{sc/ss})^{n_{e/o}}}
\eea where in the second line we have made use of the geometric series formula. 

With these assumptions, the fidelity constraint Eq. \ref{fidelitycalc} can be numerically solved for any threshold fidelity,\footnote{There is a subtlety here; unlike the fidelity for a single qubit protocol which is typically an average over unmonitored processes, here the client knows precisely how many attempts were made. Thus the fidelity here is interpretted as a minimum average quantum state fidelity, and not as an average (over the number of attempts) average quantum state fidelity.} which we will do in the next section (via postselection of the simulation). This is necessary for solving for the full rates given in Eqs. \ref{Rates2M:t}-\ref{Rates2M:m}. However, for finding the multiplexing advantage we do not need to find the full form of the rate, just its ratio with the un-multiplexed rate. 

Taking the low-success limit such that $P_{sc} \ll 1/n_{e/o}$, but taking $P_{ss}$ large such that $1-(1-P_{ss})^{n_e}$ is close to unity, we can divide Eqs. \ref{Rates2M:t}-\ref{Rates2M:m} by Eq. \ref{Rates2M:t} with $M=1$ (the un-multiplexed case) to yield the $s=2$ multiplexing improvement factors for our two strategies 

\bea\label{factors2Ma}
m_{s=2,\sigma=t} &\lessapprox& M \\
m_{s=2,\sigma=m} &\lessapprox& M (M-1) n_o n_e /n_e' \label{factors2Mb}
\eea with $n_e'$ the un-multiplexed value of $n_e$ given by the fidelity constraint. In the limit where the coherence times are very different $\tau_{co} \gg \tau_{ce}$, the cutoffs are approximately the same $n_e' \approx n_e$, and we see $m_{s=2,\sigma=m} \lessapprox M^2 n_o$ in the large-$M$ limit. 

\begin{figure*}[t]
     \centering
     \begin{subfigure}[b]{0.4
    \textwidth}
         \centering
         \includegraphics[width=\textwidth]{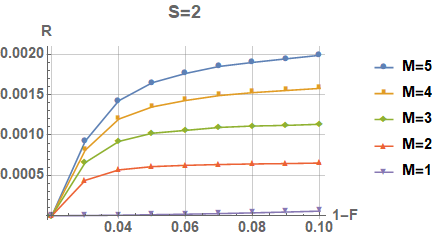}
         \caption{}
         \label{fig: multiplexing_r_vs_f}
     \end{subfigure}
     \hfill
     \begin{subfigure}[b]{0.5
    \textwidth}
         \centering
         \includegraphics[width=\textwidth]{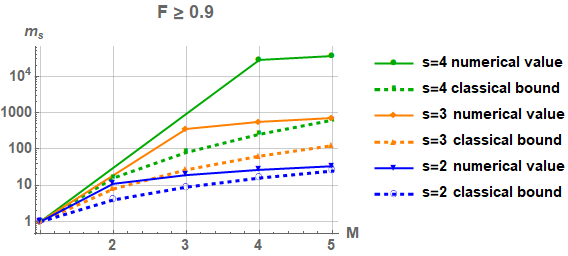}
         \caption{}
         \label{fig: multiplexing_ms_vs_M}
     \end{subfigure}
    \caption{Numerical results using the multi-server multiplexing strategy $\sigma = m$ for distant links for the client-server connection ($\eta_c = 10^{-3}$) and short links for the server-server connections $\eta_s = 10^{-1}$. Coherence times with and without EG/RSP are $20$ ms and $2.8$ s, respectively, with the full parameter values given in Table \ref{tab:appC_tab1} of Appendix D. (a) Rate-fidelity tradeoff for a scenario with $s=2$ and various numbers of servers, with rate measured in units of the duration of an EG/RSP attempt $\tau_e = 300$ ns. Note that for lower fidelities the tradeoff saturates; finite memory lifetimes effect the tradeoff by linking reduced single-qubit rates (which effect the multi-qubit rates) to a reduced the achievable final state fidelity. (b) Comparison between the $m_s$ obtained for the cases $s=2$ through $s=4$ with their corresponding classical bounds $M^s$. We included the point $m_s(M=1)=1$ for visual reference, and otherwise only consider $M\geq s$ where the multiplexing strategy we define is unambiguously specified. Note the logarithmic scaling; even for modest values of $s$ and $M$, multi-server multiplexing results in several orders of magnitude improvement in protocol execution rate. }
\end{figure*}

Crucially, the $\sigma = m$ multiplexing strategy surpasses the classical limit by a constant factor $n_o$ while exhibiting the same scaling, while the $\sigma = t$ try-and-commit strategy does not even come close to the classical limit. This is due to the $\sigma = m$ strategy's ability to prepare the second qubit faster despite more intermediate steps: there are more nodes to attempt with and the inter-node connections are assumed to be fast compared to the client-node connections, corresponding to a dense hub of nodes. Even in this toy model we can see that the choice of multiplexing strategy can dramatically effect the scaling with $M$. Also note that, while the scaling seen in Eq. \ref{factors2Ma} and Eq. \ref{factors2Mb} are the same as in the classical limit in Eq. \ref{Limit2Eq}, the constant prefactor $n_o$ can be quite large in color centers, where coherence of the order of a minute have been observed when the node is idle \cite{Bradley2019}. 

We also note that while the $\sigma=m$ multiplexing strategy gives better performance for the client, it demands more network resources at any given moment; any implementation of this protocol in a network must consider the needs of all users using a global network metric such as quantum network utility \cite{Vardoyan2023,Prielinger2025}.

\subsection{Toy Model for $s\geq2$: Numerical Results}\label{sec.numeric_multiplexing}

Although there aren't easily derivable analytic solutions for $s>2$ due to the rapidly non-trivial structure of the window problem, we can also solve this toy model numerically for $s\geq2$ by optimizing over experimental parameters. In our numerical simulations we use parameters that might arise in situations where the client and the servers (assumed to be color centers) are separated by relatively long distances ($\approx 150$ km), while the distances between servers are moderately short ($<50$ km). Further details on the numerical simulations are given in appendix \ref{app:numeric_simulations}.

In Fig. \ref{fig: multiplexing_r_vs_f} we study the $\sigma=m$ multiplexing strategy introduced in the previous section, and explore the tradeoff between rate and fidelity for the $s=2$ case for $|\alpha|^2 \leq 1/2$, as the risk of a malicious server determining the RSP qubits' states increases with the mean photon number $|\alpha|^2$ \cite{Garnier2024}. We observe that higher fidelities decrease the achievable rate as expected, and the very clear improvement in rate due to increasing the number of servers.

While a full quantum network simulation (e.g. in NetSquid \cite{coopmans2021netsquid}) would be required to reach high qubit numbers,\footnote{This is because higher qubit number protocols succeed more rarely, and are harder to directly sample.} we can study the improvement that the multiplexing strategy ($\sigma=m$) achieves for the preparation of two, three and four qubits $s\in\{2,3,4\}$ using our simple toy model, as shown in Fig. \ref{fig: multiplexing_ms_vs_M}, using the toy model numerical simulation given in appendix \ref{app:numeric_simulations}. In particular we focus our study on the cases with $s\geq M$, since these cases take the most full advantage of the two different coherence times $\tau_{co}$ and $\tau_{co}$. We observed that the improvement surpasses the classical limit of $M^s$ in all cases. Note that the gap between $m_s$ and the classical bound $M^s$ increases with $s$ for the same parameter regime. The reason for this is threefold: firstly, a single sever (no-multiplexing) implementation stores each qubits for larger amounts of time compared to the multiplexing strategy ($P_{sc} < P_{ss}$). Secondly, more qubits means less storage time-per-qubit is permitted for a fixed, $s$-independent final state fidelity $F$, effectively reducing the time qubits can be held in storage for. Lastly, and more abstractly in the language of the window problem \cite{Davies2024}, given a fixed window $w$, increasing the number of required qubits $s$ yields a stronger suppression of the rate achievable due to the finite window; the protocol feels the effects of the window more strongly as $s$ increases to its maximum theoretical value $s=w$, beyond which the protocol is ill defined.




We conclude by noting that we have assumed a highly symmetric network protocol and configuration, such that server and client keep their parameters fixed ($P_{sc}$, $P_{ss}$), throughout the whole protocol. Future work on realistic networks could use machine learning to characterize realistic networks, which will lack the nice symmetries that we exploited in this analysis, as well as adaptively adjusting parameters based on network performance \emph{en vivo}.  


\section{Summary and Outlook}

In this work, we have identified classical limits to the improvement to a quantum protocol induced by multiplexing, and introduced two beyond-classical methods for surpassing them: single click quantum multiplexing and multi-server multiplexing. We have studied the performance of these new methods by analyzing rate-fidelity tradeoffs and scalings with the multiplexing resource $M$ for simple toy models. Here, we found that single click quantum multiplexing improves performance by enabling asymmetric resource utilization and $>2$-party links, and that multi-server multiplexing similarly improves performance by making use of many inter-linked servers (or one server with multiple communication qubits) to provide service to a single client, especially for many-qubit protocols. In both cases, these methods improve the rate-fidelity tradeoff (ubiquitous across almost all domains quantum science \cite{Banaszek2001,DAriano2009,MooreTibbetts2012,Almeida2018,MeyerScott2020,Tanji2024}\footnote{Though interestingly, not for single photon detection, where the dark count-detection rate tradeoff can---at least in principle---be circumvented through detector design; instead, the rate of detection impacts which single photon temporal wavepackets can be efficiently detected \cite{Propp2020}.}) and can yield an improvement to the quantum bit or protocol execution rate upper-bounded by limits that we identify. In practice, to achieve the highest rate for a quantum network protocol one should make use of all available multiplexing techniques: semiclassical, quantum, and multi-server. This is especially important for measurement-based blind quantum computation where there is a very large overhead in the number of qubits needed \cite{Broadbent2009}. Here, multi-server multiplexing especially can improve the unfavorable scaling due to the many high-fidelity qubits that are required for noise robust verification. 

Furthermore, we have seen in our simple toy model for multi-server multiplexing how different strategies give rise to rate improvements $m_s$ for symmetric networks. We hope this will motivate further work studying next-generation multiplexing techniques and multiplexing strategies in quantum networks e.g. in NetSquid \cite{coopmans2021netsquid}, especially for networks with high degrees of asymmetry \cite{Prielinger2025}, employing adaptive strategies \cite{Tacettin2025}, or with pre-existing entanglement \cite{Iesta2023}. These, especially in the large-$s$ limit, would form of the basis of experiments utilizing single click quantum multiplexing and multi-server multiplexing in long-distance in deployed quantum networks \cite{Pompili2022,Stolk2025}.

In this paper, we have shown how our next-generation multiplexing techniques can improve both the single qubit rates by a factor $m$, and the $s$-qubit rates by a factor $m_s$. In all cases, this improvement is due not to improved hardware quality at all, but due to \emph{more} hardware. For a quantum internet network, networking lower-quality quantum servers improves their collective performance. This is especially appealing due to recent developments in software stack development (including deployable local operating systems for quantum nodes \cite{DelleDonne2025}) and manufacturing capabilities \cite{Xue2021,Boter2022}, which pave the way towards scaling up quantum device production and deployment. Augmented by multiplexing, these enable a possible route to transition from a prototype dial-up style quantum internet (slow, unreliable service, limited application) to a widely-applicable broadband style quantum internet (fast, multiplexed, wide application), all utilizing the same underlying components currently in active development. 

Furthermore, the non-trivial relationship between the behaviors of $m$ and $m_s$ with $M$ is indicative of a broader aspect of quantum network theory. In a classical network protocol the executes-per-second of a protocol and the bits-per-second of a link are usually directly connected; in general a linear speedup in one corresponds to a linear speedup in the other. However, the quantum bits-per-second and the quantum executes-per-second may show very different behaviors with $M$; here, we have seen that the quantum bits-per-second tend towards an at-most linear speedup with $M$ (i.e. $m\leq M$), whereas the quantum executes-per-second may have a superlinear speedup (i.e. $m_s \leq M^s$ for $s\ll w$). This is due to the relative fragility of quantum information in comparison to classical information \cite{Terhal2012}; only in the infinite-window limit (i.e. the perfect memory limit) do we recover a classical relationship between the quantum bits and executes-per-second. Only when decoherence is removed---either by perfect memories or by quantum error corrections---does the scaling become linear regardless of the number of qubits in the protocol $m_s\sim M$, as is the case classically. We hope that this non-trivial relationship between qubit rates and quantum execution rates on near-term quantum devices will continue to be studied in future works. 
 
Here we have identified semiclassical limits to the multiplexing advantage, and shown how to exploit quantum mechanics to break them. However and emphatically, we have not here identified fundamental Heisenberg limits to the multiplexing advantage, nor do we claim our quantum and multi-sever multiplexing methods are optimal in a quantum sense. One can readily imagine more advanced techniques for quantum multiplexing that, following the ideas of the original quantum multiplexing proposal in Ref. \cite{LoPiparo2019} or experimental efforts to coherently manipulate higher-dimensional degrees of freedom of photons \cite{Christ2012,Ansari2018,Ansari2018b,MeyerScott2020}, encode qubits across many pulse in higher dimensional qudits across servers,\footnote{Indeed, even basic questions like ``How many qubits can one encode in an $N$-pulse sequence?" are non-trivial to answer except in the most basic cases i.e. if we limit to the zero and one photon-per-pulse subspace the answer is $N$ if we assume overall phase stability, which is achieved using the single click protocol \cite{vanDam2025}. However, while relaxing phase stabilization should result in $N-1$ qubits since the overall phase constitutes a classical gauge degree of freedom \cite{Bartlett2007}, as of yet there is no protocol for extracting all of the qubits with high probability and the practical bound may be much lower.} or multi server multiplexing strategies utilizing asymmetric adaptive strategies and advanced memory configurations to further improve network performance. Assessing the optimality of those techniques will require identifying the fundamental limits to multiplexing advantage. It is well known that remote state preparation is a less demanding task than quantum state teleportation, with the latter requiring entanglement between two parties. Yet the semiclassical limits we've derived for the multiplexing advantage are identical for both RSP and EG processes; it is an open question whether Heisenberg-like quantum limits to multiplexing will reflect this fundamental difference between RSP and EG. We hope that in addition to guiding experimental design, this work will inspire theoretical development of even better next-generation quantum and multi-server multiplexing techniques, as well as inspire researchers to continue the search for the true quantum limits to multiplexing.

\section{Acknowledgements}

We gratefully acknowledge assistance provided by Bethany Davies and Thomas Beauchamp with our analysis. We are also grateful to Mariagrazia Iuliano, Bart van der Vecht, \'{A}lvaro Iñesta, Harold Ollivier, Maxime Garnier, Tracy Northrup, Ronald Hanson, and Benedikt Tissot for stimulating and illuminating discussion throughout the course of this project, as well as the membership of the Quantum Internet Alliance for their broad support.

This project has received funding from the European Union’s Horizon Europe research and innovation program under grant agreement No. 101102140.


\bibliography{Multiplexing}

\appendix

\section{Derivation of Semiclassical Limit to Multiplexing} \label{app:derive}

Here we derive rigorously the semiclassical upper-bounds to the multiplexing advantage introduced in section II. Throughout our paper we have defined the multiplexing advantage $m_s$, Mathematically, we can define $m_s$ through the expression

\bea\label{Physical}
R_{s|F}(M)=m_s R_{s|F}(1). \eea

If we are in the regime where almost all instantiations of the protocol fail, then each implementation may be considered a Poisson or geometric process. This enables us to interpret $m_s$ physically, in this limit, as the number of un-multiplexed implementation of the $s$-qubit quantum protocols that would need to run in parallel to achieve the same rate as a single instantiation of the $M$-fold multiplexed protocol. 

We now begin with the derivation of the multiplexing advantage to the single-qubit rate to prepare or transmit a single qubit of fidelity $F$, which we provided a semiclassical limit for in Eq. \ref{Limit1Eq}. Rearranging Eq. \ref{Physical} for $s=1$, we find 

\bea\label{singlequbitmdef}
m = \frac{R_{p_M|F}(M)}{R_{p|F}(1)}.
\eea Here $p$ is the probability that an un-multiplexed attempt to generate entanglement or remotely prepare a state succeeds, and $p_M$ is the probability that each attempt of the multiplexed attempts succeed. We can take $p_M = p$ to give an upper limit to $m$, which we assume henceforth. To reiterate, we also assume that the classical multiplexing does not negatively impact the maximum achievable fidelity for the protocol, as is the case e.g. for temporal multiplexing, up to a practical limit imposed by the physics e.g. how many pulses can fit in a beamline without overlap, how many spatial or spectral modes are accessible for multiplexing.

We now recast our un-multiplexed rate $R_{p|F}(1)$ in terms of expectation values $\tau_p$: the time to achieve the first (and here, only) success, which is a random variable. We assume this to be a Poisson process as is done in Ref. \cite{Davies2024}. In this way for the unmultiplexed protocol, we can write 

\bea\label{unmultiplexed rate}
R_{p|F}(1) = \frac{1}{\mathbb{E}(\tau_p) (\tau_1+\tau_c)}\eea with $\tau_1$ the duration of a single attempt's propogation time and $\tau_c$ the duration of any additional classical communication needed to complete the attempt.  

For the multiplexed case $R_{p|F}(M)$, the behavior of 
the random variable $\tau_p$ will be affected by the exact structure of the multiplexing e.g. exact simultaneity as in spectral mode multiplexing vs sequential as in temporal mode multiplexing. However, we can upper bound the multiplexed rate by the simultaneous model where all $M$ batched attempts occur at the same (first) time step, and where a batch succeeds if any non-zero number of attempts succeed. Thus,  $R_{p|F}(M) \leq \frac{1}{\mathbb{E}(\tau_{1-(1-p)^M}) (\tau_M+\tau_c')}$, with $\tau_M$ the time needed for the $M$ pulses to propagate and $\tau_c'$ any additional time needed for a shared communication round for the batch. Thus we calculate the upper bound of the ratio of the rates

\bea\label{singlequbitlimit1}
m \leq \frac{\mathbb{E}(\tau_{p})}{\mathbb{E}(\tau_{1-(1-p)^M})}\frac{\tau_1 +\tau_c}{\tau_M + \tau_c'}.
\eea  Since $\mathbb{E}(\tau_{p})=1/p\,\forall p$, we find  

\bea\label{singlequbitlimit2}
m \leq \frac{1-(1-p)^M}{p}\frac{\tau_1 +\tau_c}{\tau_M + \tau_c'} \leq M \frac{\tau_1 +\tau_c}{\tau_M + \tau_c'} \leq M \nonumber \\
\eea as desired. Equality with the first upper bound is achieved in the limit $p\rightarrow 0$, and equality with the second upper bound is achieved when the time $\tau_M+\tau_c'$ to perform a batch of $M$ multiplexed attempts takes an equal duration to performing a single un-multiplexed attempt $\tau_1+\tau_c$ (i.e. exact simultaneity). 

To derive our second limit Eq. \ref{Limit2} for $s$-qubit protocols is more involved, and requires using techniques and results derived in Ref. \cite{Davies2024}. There, the subject was a general study of the problem of finding the average number of attempts needed to achieve $s$ successful executions of a protocol on a quantum network within $w$ attempts, colloquially known as the ``window problem" of scan statistics \cite{Glaz2001}. The primary quantity of interest is the expectation time $\mathbb{E}(\tau_{w,s,p})$ to get $s$ successes within a window of $w$ attempts, where each attempt succeeds with probability $p$. We begin again with our definition of the multiplexing advantage, now for an $s$-qubit protocol 

\bea\label{squbitmdef}
m_s = \frac{R_{s,p_M|F}(M)}{R_{s,p|F}(1)}.
\eea To translate this to a window problem, we assume that generation of a state with a given minimum fidelity $F$ is equivalent to generating a state within a specified window $w$ as discussed in the main text. Notably, this assumption of a single fixed window is the key assumption we exploit in studying multi-server multiplexing. To give an upperbound, we also utilize that the time between successes is minimized when the $M$ multiplexed attempts are equally spaced in time  (as opposed to so-called ``bursty windows" \cite{Braverman2012}). Intuitively, this lower bounds the expected time to success by minimizing gaps between attempts with each attempt being equally likely to be successful, and allowing for successes to occur \emph{within} our original attempt time $\tau_1 + \tau_c$, removing the dead communication time from the un-multiplexed case. 

By defining the duration of each of the $M$ unbatched multiplexed attempts to be  $\tau_{1_M}=(\tau_1 + \tau_c)/M$ corresponding to this idealized protocol with no wasted communication time, we arrive at an initial upper bound for the multiplicative improvement factor

\bea\label{mqubitlimit1}
m_s \leq \frac{\mathbb{E}(\tau_{w,s,p})}{\mathbb{E}(\tau_{Mw,s,p})}\frac{\tau_1 +\tau_c}{\frac{\tau_{1}+\tau_c}{M}} = M\frac{\mathbb{E}(\tau_{w,s,p})}{\mathbb{E}(\tau_{Mw,s,p})} \equiv m_s^*.\nonumber \\
\eea In this upper limit, we directly observe the effect of multiplexing as increasing the number of attempts we make (i.e. the window $w$) by a factor of $M$ while simultaneously reducing the duration of each attempt by a factor of $M$. Note that for $s=1$ the dependency on window itself disapears from $\mathbb{E}(\tau_{w,s,p})$ and we recover the single-qubit upper limit Eq. \ref{singlequbitlimit2}: $m_{s=1}^*=M$.

The task now is to upper bound $m_s^*$ itself for $s>1$. To do this we first consider several limits where the behavior of $m_s^*$ is easy to describe.  

Consider the low-success limit $p\rightarrow0$. Here, $\mathbb{E}(\tau_{w,s,p}) \sim \frac{1}{{w-1 \choose s-1}p^s}$ so that

\bea\label{nowproblim}
\lim_{p\rightarrow 0} m_s^* = M\frac{{Mw-1 \choose s-1}}{{w-1 \choose s-1}}.\eea Note that in the limit $w\gg s$, this is upper-bounded $m_s^*\leq M^s$ with equality only in the large $w$-limit. 

Consider also the opposite limit $p\rightarrow 1$, where $\mathbb{E}(\tau_{w,s,p})$ becomes approximately independent of the window size, so that $\lim_{p\rightarrow 1} m_s^* = M $, which we note is strictly less than the opposite $p\rightarrow 0$ limit for $M>1$.

As a final limit, consider the infinite window limit $w=\infty$, where again $\mathbb{E}(\tau_{w,s,p})$ becomes independent of the window size, so that $\lim_{w\rightarrow \infty} m_s^* = M $.

Now, we make use of several properties of the expectation values $\mathbb{E}(\tau_{w,s,p})$. The first as discussed in Ref. \cite{Davies2024} is that $\mathbb{E}(\tau_{w',s,p}) \geq \mathbb{E}(\tau_{w,s,p}) $ for $w'>w$, with equality only at $w=\infty$ or $p=1$. The second is a physically motivated and numerically verified assumption: that the first derivative of the expectation values are monotonic with window size i.e.

\bea\label{assumption}
0\geq \frac{d}{dp} \mathbb{E}(\tau_{w',s,p}) \geq \frac{d}{dp} \mathbb{E}(\tau_{w,s,p}), \,\,w'>w
\eea with equality only asymptotically as $w=\infty$ or $p=1$. This assumption aligns with physical intuition; the larger the window $w$, the better the behavior of $\mathbb{E}(\tau_{w,s,p})$ is described by its asymptotic behavior in the infinite-window limit \cite{Davies2024}, which is the regime is where the expectation time $\mathbb{E}(\tau_{w,s,p})$ is \emph{least} sensitive to changes in the success probability $p$, scaling linearly $\mathbb{E}(\tau_{w,s,p})\propto p$. More physically, as the \emph{quality} of physical memory qubits improves, we expect decreases in individual success probabilities have less of an impact on the $s$-qubit expectation time $\mathbb{E}(\tau_{w,s,p})$. If this assumption were not to hold for some system, it would mean that, for networks using those systems, improving the memory quality would increase the sensitivity of the protocol performance to the effects of decoherence, which seems unphysical. 

We now consider what it would mean for $m_s^*$ to be non-monotonic in $p$ for finite $w$: namely, that there is at least one non-boundary value of $p$ such that $\frac{dm_s^*}{dp}=0$. From its definition Eq. \ref{singlequbitlimit1}, this requires

\bea\label{derivationeq1}
\!\!\!\!\!\!\!\!\!\!0 = \left(\frac{d}{dp}\mathbb{E}(\tau_{w,s,p})\right) \mathbb{E}(\tau_{Mw,s,p})- \mathbb{E}(\tau_{w,s,p})\left(\frac{d}{dp}\mathbb{E}(\tau_{Mw,s,p})\right).\nonumber \\ 
\eea Rearranging, we find 

\bea\label{derivationeq2}
\frac{\frac{d}{dp}\mathbb{E}(\tau_{Mw,s,p})}{\frac{d}{dp}\mathbb{E}(\tau_{w,s,p})} = \frac{\mathbb{E}(\tau_{Mw,s,p})}{\mathbb{E}(\tau_{w,s,p})}. \nonumber \\ \eea However, we know $\mathbb{E}(\tau_{Mw,s,p})\leq \mathbb{E}(\tau_{w,s,p})$ and have argued for $\frac{d}{dp} \mathbb{E}(\tau_{Mw,s,p}) \geq \frac{d}{dp} \mathbb{E}(\tau_{w,s,p})$, with equalities only at $p=1$ or $w=\infty$. Thus, since the left hand side is greater than unity and the right hand side is less than unity except at $p=1$ or $w=\infty$, we conclude that for finite $w$ $m_s^*$ is strictly monotonic in $p$. 

Since the largest value of $m_s^*$ is observed as $p\rightarrow0$, and $m_s\leq m_s^*$, we arrive at our limits Eq. \ref{Limit2Eq} and Eq. \ref{Limit2Eq2}:

\bea\label{finalslimit}
m_s \leq M\frac{{Mw-1 \choose s-1}}{{w-1 \choose s-1}} \stackrel{w\gg s}{\rightarrow} \leq M^s.\eea

Lastly, we note that for any physical implementation of an experiment using classical multiplexing there is limit to how much information may be encoded in a channel so that $M\leq M_c$, as is well documented in the literature \cite{Nyquist1928,Shannon1949,Caves1994,Adami1997,Schumacher1997,Lloyd1997,Wilde2017}.

\onecolumngrid

\section{Multiplexing RSP details} \label{app:RSP appendix}
After the transformation in Eq. \eqref{eq:dm_rsp}, we obtain the full density matrix 
\begin{align}
    \rho_{RSP|k+}=&\frac{e^{-M\eta_c|\alpha|^2}}{\pqty{1-e^{-\eta_c|\alpha|^2/2}}\pqty{1-\eta_s\xi^2}+\xi^2\frac{\eta_s}{2M}\pqty{1+\frac{\eta_c|\alpha|^2}{2}}}\bigg\{
    \pqty{1-e^{-\eta_c|\alpha|^2/2}}\pqty{1-\xi^2}\ketbra{0}\nonumber\\
    &+\frac{\sqrt{\eta_c\frac{\eta_s}{M}(1-\xi^2)}}{2}\abs{\alpha}\xi\pqty{e^{i\theta_k}\ketbra{1}{0}+\text{h.c.}}
    +\xi^2\bqty{\frac{
    }{2M}\pqty{1+\frac{\eta_c|\alpha|^2}{2}}+\pqty{1-\eta_s}\pqty{1-e^{-\eta_c|\alpha|^2/2}}}\ketbra{1}
    \bigg\}.
\end{align}
This yields the fidelity to the desired output state $\ket{+_{\theta_k}}=(\ket{0}+e^{i\theta_k}\ket{1})/\sqrt{2}$ to be
\begin{align}\label{eq:Fid_full}
    F= \frac{1}{2}\bqty{1+\frac{\sqrt{\eta_c\frac{\eta_s}{M}(1-\xi^2)}\abs{\alpha}\xi}{\pqty{1-e^{-\eta_c|\alpha|^2/2}}\pqty{1-\eta_s\xi^2}+\xi^2\frac{\eta_s}{2M}\pqty{1+\frac{\eta_c|\alpha|^2}{2}}}}.
\end{align}
We maximize this fidelity over the bright state population $\xi$, to obtain the optimal value 

\begin{align}
    \xi^2=\frac{\pqty{1-e^{-\eta_c|\alpha|^2/2}}}{
    2\pqty{1-e^{-\eta_c|\alpha|^2/2}}+\eta_s\bqty{\frac{1+\eta_c|\alpha|^2/2}{2M}-\pqty{1-e^{-\eta_c|\alpha|^2/2}}}} 
    \sim
    \frac{M\eta_c}{\eta_s}|\alpha|^2.
\end{align}
where the last form given is the small-$\alpha$ limit. Using the full description in Eq. \eqref{eq:Fid_full} yields Eq. \eqref{eq:Fid_Rsp}.

The probability to get a successful detection event for any single client device is
\begin{align}\label{Rell}
    \Pr[\mathrm{single\,channel}] = \frac{4e^{-(M-\frac{1}{2})\eta_c|\alpha|^2}(1-e^{-\eta_c|\alpha|^2/2})\bqty{4M(1-\eta_s)\pqty{1-e^{-\eta_c|\alpha|^2/2}}+2\eta_s\pqty{1+\frac{\eta_c|\alpha|^2}{2}}}}{4M(2-\eta_s)\pqty{1-e^{-\eta_c|\alpha|^2/2}}+2\eta_s\pqty{1+\frac{\eta_c|\alpha|^2}{2}}} \sim 2\eta_c|\alpha|^2,
\end{align}
As in the previous analysis, we define the time of each attempt $\tau_e$ to be unity so that the single client device success probability is the same as the rate for a specific client device $R_k$. Multiplying this number by $M$ yields Eq. \eqref{eq:Rate_Rsp}, as this is the success probability/rate for any one of the channels to succeed.

\subsection{Continuous demand}
We will now consider two separate cases not studied in the main text but with very similar results that may be of particular interest to readers with an eye towards quantum network protocols and resource allocation. Wheras in the maintext we assume all photonic client devices to be held by a single user, here we will assume they are held by individual users who will either continuously demand RSP qubits on the server, or be content with a single RSP qubit and leave the protocol after obtaining it e.g. for a multiparty secure computation \cite{Polacchi2023}. In the next subsection we will consider that the clients only demand one qubit, and after obtaining this, they leave the queue.

We will in this section explicitly write the dependence on $M$ in the rate, fidelity, and multiplexing improvement for clarity, i.e. $R_k(M)$, $F(M)$, and $m(M)$. 
Using the multiplexed protocol for $M$ clients the rate and fidelity are given by $R_k(M)$ and $F(M)$, whereas, the rate and fidelity for the non-multiplexed case is $R_k(1)/M$ and $F(1)$.

\begin{figure*}
     \centering
     \includegraphics[width=\textwidth]{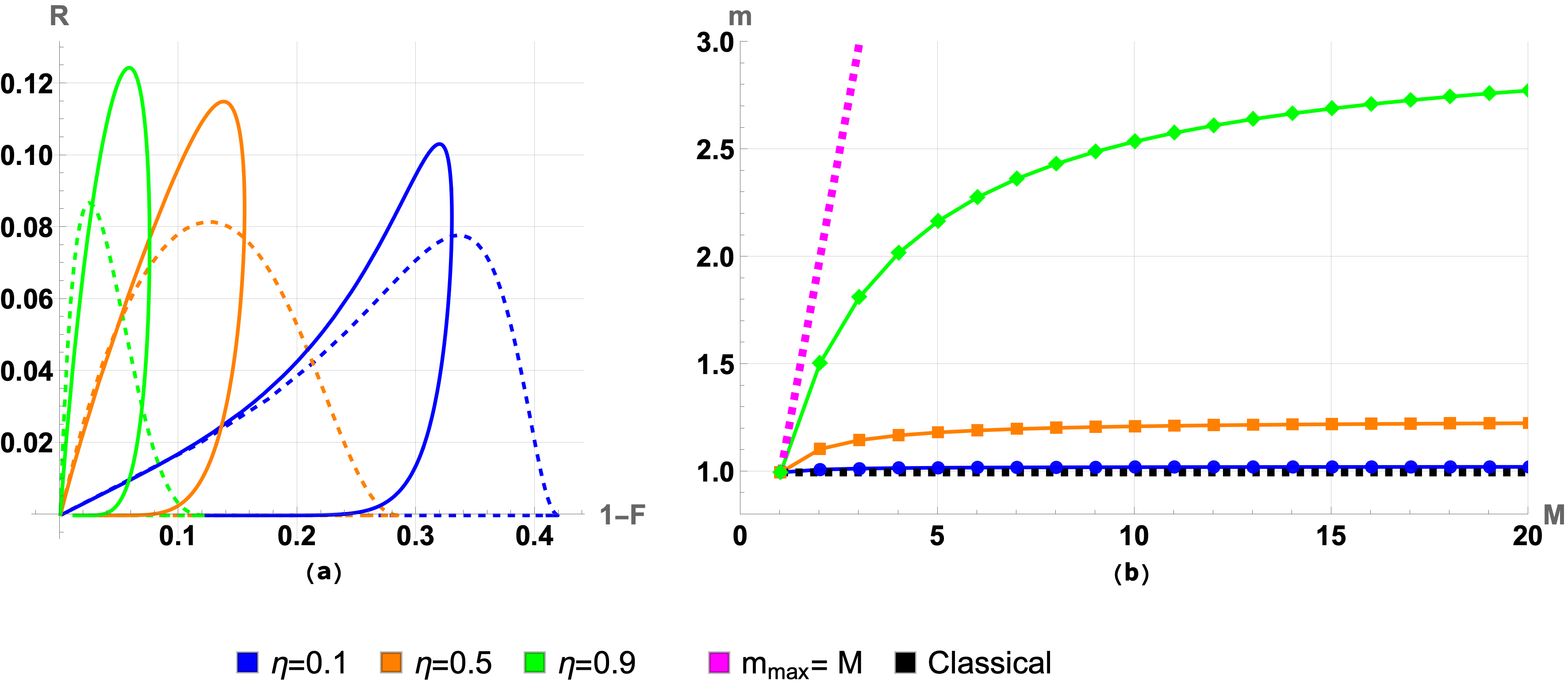}
    \caption{Single click quantum multiplexing between several client devices held by different clients and a single node in a quantum network with continuous RSP demand. (a) Fidelity-rate curves parameterized by $\gamma=\eta_c |\alpha|^2$. Here, $M=5$ clients each with their own photonic client device continuously compete for access to a single processing node, with both time sharing (solid) and single click quantum multiplexing (dashed) strategies. (b) Multiplexing gain in this case $m$ as function of $M$ for RSP with fidelity $F_\mathrm{min}=1-10^{-3}$ using $M$ competing clients. The theoretical maximum with a perfect server and the classical bound is plotted using the magenta and red black lines, respectively.}\label{fig: fidelity rate curves for clients 2}
\end{figure*}

The rate versus fidelity is plotted in Fig. \ref{fig: fidelity rate curves for clients 2}a. We find in the high-fidelity regime that 
\bea\label{mfactorclient2}
m(M) \sim \frac{M (1 - \eta_s +\frac{\eta_s}{4})}{M(1-\eta_s) + \frac{\eta_s}{4}}  
\rightarrow 1+\frac{\eta_s}{4(1-\eta_s)},
\eea
where again in the last line we consider taking the limit $M\rightarrow\infty$. We plot the behavior of $m$ with $M$ in Fig. \ref{fig: fidelity rate curves for clients 2}b. Again, we see that the multiplexing factor increases as $\eta_s$ approaches unity, which is also seen in Fig. \ref{fig:use-case client}. Physically, this corresponds to the single click quantum multiplexing scheme allowing for an improvement in rate when providing service to multiple clients as compared to the standard time-sharing method. In the limit of a perfect efficiency connection to the server, we find $m\to M$ for each user.

\subsection{Single qubit demand}
\begin{figure*}
     \centering
     \includegraphics[width=\textwidth]{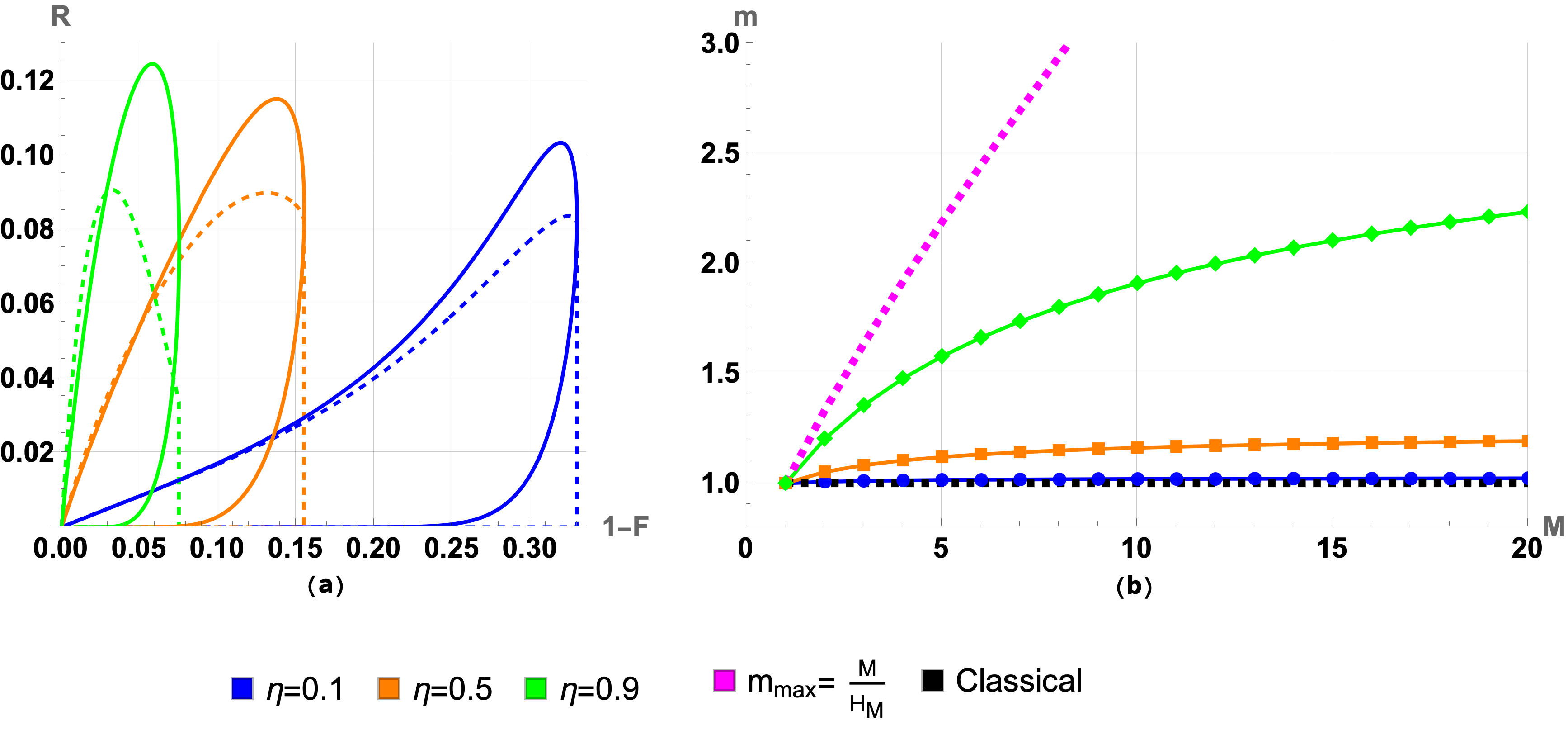}
    \caption{Single click quantum multiplexing between several client devices held by different clients and a single node in a quantum network, where after a successful RSP each client stops participating in the protocol. (a) Fidelity-rate curves parameterized by $\gamma=\eta_c |\alpha|^2$. Here, $M=5$ separate clients each with their own photonic client device compete for access to a single processing node, with both time sharing (solid) and single click quantum multiplexing (dashed) strategies. After obtaining a single qubit the client leaves the competetion. (b) Multiplexing gain in this case $m$ as function of $M$ for RSP with fidelity $F_\mathrm{min}=1-10^{-3}$ using $M$ competing clients. The theoretical maximum with a perfect server and the classical bound is plotted using the magenta and black dashed lines, respectively.}
    \label{fig: single use}
\end{figure*}

We now consider single qubit demand. After successfully preparing a qubit on the server, the client will not want to prepare any further qubits, and thus leaves. In the un-multiplexed case, the rate and fidelity are the same as in the previous case, as the time-sharing strategy work similarly. However, in the multiplexed case after each successful preparation the number of participating clients changes. Hence, the formula for the fidelity changes slightly, as it depends on the number of clients. Thus, after each successful heralding, the BSP and  $\alpha$ are updated to ensure the same target fidelity. The rate of success is given by
\begin{align}
    R = \left[\sum_{m=1}^M \frac{1}{mR_k(m)}\right]^{-1},
\end{align}
where $R_k(m)$ uses updated values for the bright state probability and $\alpha$.
With this we can make the rate-fidelity tradeoff plots and the multiplexing advantage, see Figs. \ref{fig: single use}ab. The large drop in the rate at specific fidelities is caused by, a single user no longer being able to get a lower target fidelity. This sets a lower bound on the target fidelity with this method. However, as can be seen, at this target fidelity time-sharing is more efficient for the clients than single click quantum multiplexing.  Compared to continuous demand case studied in the previous section, we see smaller multiplexing gains because, on average, fewer users are participating in the protocol at any given time. In the high-fidelity regime, we find
\begin{align}
    m(M) \sim \frac{M(1-\frac{3\eta_s}{4})}{M(1-\eta_s)+H_M\frac{\eta_s}{4}}\rightarrow 1+\frac{\eta_s}{4(1-\eta_s)},
\end{align}
where $H_m$ is the $m$'th harmonic number \cite{Ri2024}, and the limit is taken for $M\to \infty$. We can clearly see that this is smaller than the continuous demand, as $H_m>1$.  We again note that $m(M)\to M/H_M$, when $\eta_s\to1$ and $\abs{\alpha}\to 0$.



\section{Review of Single Click Entanglement Generation}\label{app:SCreview}

 Here, we briefly review the single click entanglement generation scheme in more detail assuming perfect photo-detection (unit efficiency, no dark counts), following closely the exposition given in Ref.~\cite{MoraThesis}, to better understand the physics of inter-server links. For a more complete description including the low loss limits, see Ref.~\cite{Hermans2023}
 
 The more common double click entanglement generation protocol can be implemented on Nitrogen vacancy (NV) centers using the Barrett-Kok scheme is commonly used: a spin-selective cycling transition is used such that light emitted from an NV center will also emitted in a superposition of an early time bin and a late time bin, with the state entangled with the internal state of the electron \cite{Barrett2005,Hensen2016phd}. Thus, by making use of the intermediate bell state measurement, it is possible for a remote client to prepare the state of the NV center's electron by manipulating the phase $\theta$ of the late pulse induced by the interferometer. Using a sequence of controlled rotations $CR_Y^{\frac{\pi}{2}}$ and $CR_x^{\frac{\pi}{2}}$, the state of the electron can be moved to carbon atoms in the lattice for longer-term storage. 

Single click entanglement generation works similarly, but instead of encoding quantum information into a single photon in superposition between two modes (dual rail), it encodes quantum information in the presence or absence of a photon (single rail). This requires phase stabilization, but works particularly well in the high-loss limit \cite{Campbell2008} and has been implemented in NV centers \cite{Humphreys2018}. One starts by creating the entangled superposition state between a communication qubit on a server node and an outgoing optical mode
\bea\label{singleclick}
\ket{\beta} = \xi^2 \ket{0}_{\rm server}\ket{1}_\gamma + \sqrt{1-\xi^2} \ket{1}_{\rm server}\ket{0}_\gamma,
\eea with the first ket referring to the internal state of the server qubit and the second ket reffering to the state of the optical field ($0/1$ denotes the absence/presence of a photon). Experimentally, $\xi^2$ is bright state population of an NV center, and can be tuned to a desired value. 

Creating such a state as described by Eq. \ref{singleclick} at two server nodes, interfering the resulting lightfields on a beam splitter, and measuring the outputs results in an entangling measurement between the two nodes. In the absence of dark counts, the resulting state is (dropping subscripts)

\bea\label{singleclickstate}
\hat{\rho}_{\rm SC} \approx F_{\rm SC} \ket{\Psi_{\pm}}\bra{\Psi_{\pm}} + (1- F_{\rm SC}) \ket{00}\bra{00}.
\eea Here, $\ket{\Psi_{\pm}}$ are maximally entangled Bell states with $\pm$ determined by which detector clicks, $\ket{00}$ is a classically correlated (but useless) state where both qubits are in the computational $0$ state, and $F_{\rm SC}$ is the Bell State fidelity. In the high-loss regime \cite{Hermans2023}, this is given by

\bea\label{SingleClickFidelity}
F_{\rm SC} = \frac{1}{2} (1-\xi^2) (1+\sqrt{V})(1-p_{\rm ph})
\eea with $V$ the fringe visibility of a Hong-Ou-Mandel experiment between the two light fields, and $p_{\rm ph}$ a dephasing parameter reducing the quality of the state produced due to dephasing during the (classical) communication time. Note that to first order the fidelity $F_{SC}$ is independent of loss in the high-loss limit (for details, see Ref.~\cite{Hermans2023} as well as Section 3.4 of Ref.~\cite{MoraThesis}). In our analysis in this paper, we have taken $V=1$ (perfect mode-matching) and $p_{\rm ph}=0$ (no dephasing of the communication qubit during entanglement generation), so that $F_{SC} = 1-\xi^2$. Since we are interested in using this entanglement to teleport quantum states, we use the results of Horodecki, Horodecki, and Horodecki \cite{Horodecki1999} that 
\bea\label{teleportfidelity}
F = \frac{2F_{SC} + 1}{3} = 1 - \frac{2\xi^2}{3}. 
\eea

We also consider the rate attainable with single click entanglement generation. Defining the (symmetric) single-photon loss due to transmission to be $\eta$, we find that, given a successful initialization of the state Eq. \ref{singleclick}, the probability of the single click EG protocol succeeding in an attempt is 

\bea\label{SingleClickProbability}
P_{\rm SC} &=& 2\eta (1-\eta)(1-\xi^2)\xi^2 \nonumber\\
&\approx& 2\eta \xi^2,
\eea where in the second line we've assumed $\eta\ll 1$. Here the factor of $2$ coming from the fact that a photon coming from either side results in a success.

Immediately we observe a rate-fidelity trade-off, a higher value of $\xi^2$ means a higher percentage of successful entanglement generation attempts, but the quality of the entanglement generated is reduced proportionally. 

\section{Numerical simulations} \label{app:numeric_simulations}

Here we give further details about the method used to estimate the rates in scenarios that use $M$ nodes to prepare $s$-qubits using multi-server multiplexing (Figures \ref{fig: multiplexing_r_vs_f} and \ref{fig: multiplexing_ms_vs_M}). The problem is certainly complex, and conversely in our simulations we did not use the full parameter space available to maximize the rates for our toy model. However, our simulations suffice to study the improvement that the multiplexing strategy achieves, and to demonstrate that this advantage is larger than the classical limit $M^s$. For this purpose we used the following parameters for the transmission losses, decoherence times, and maximum number of attempts (Tab. \ref{tab:appC_tab1}) 
\begin{table}[hb]
    \centering
    \begin{tabular}{c|c}
        \text{Parameter} & \text{Value} \\
        \hline
        $\eta_{s}$ & $10^{-1}$\\
        $\eta_{cs}$ & $10^{-3}$\\
        $\tau_{e}$ & 300 \text{ns}\\
        $\tau_{co}$ & 2.8 \text{s}\\
        $\tau_{ce}$ & 20 \text{ms}\\
        $n_{e}$ & $10^3$ \\
        $F_{min}$ & $0.9$ \\
    \end{tabular}
    \caption{List of parameters used in the numerical simulation and their respective values.}
    \label{tab:appC_tab1}
\end{table}

We restricted ourselves to cases where the connection between the client and all available servers are identical, so that they have the same transmission efficiency $\eta_{c}$ (spatial symmetry). Similarly, the links between servers are all assumed to have the same efficiency$\eta_s$. Moreover, we also consider that the client (server) uses a fixed $|\alpha|^2$($\xi$) for RSP (EG) across all rounds of the protocol (temporal symmetry). We used multi-server multiplexing between server and client based on single click RSP \cite{vanDam2025}, see Tab.\ref{tableprobs}, as well as a target minimum fidelity of $0.9$. We assumed servers to be connected by double click EG and do not consider quantum multiplexed connections between servers.\footnote{Here we did not include quantum multiplexing between nodes, but one certainly could do so and study the combined effect of quantum and multi-server multiplexing advantages.} Additionally, we restricted ourselves to the low-amplitude limit such that $|\alpha|\leq \sqrt{ 1/2}$, motivated physically by a security requirement as the probability a malicious server could determine the state of the RSP qubits grows with the mean photon number $|\alpha|^2$ \cite{Garnier2024} and alignment with the mean photon number (after loss by $\eta_c = 10^{-3}$) of a recent WCP RSP experiment \cite{Iuliano2024}, as well as prior quantum key distribution experiments \cite{Boaron2018}.


Modeling the idealized server-server connection poses a question of how idealized to be, as in the no-imperfections limit the fidelity of the Barret-Kok double click entanglement generation scheme is unity and the success probability is $P=\eta^2/2$ for deterministic quantum emitters \cite{Barrett2005,Ji2022}. However, implementations in NV centers remain far below these optimsitic values even at short distances \cite{Bernien2013}, and have intrinsically limited fidelity. Other color centers claim better performance e.g. $P=0.44$ and $F=1-10^{-3}$ for Silicon \cite{photonicinc}, which we use as optimistic target values for our simulation. 

To calculate the figures presented in Sec. \ref{sec.numeric_multiplexing}, we made use of a sampler. This means that for a total number of rounds $N$, we draw from geometric distributions the number of attempts that takes to successfully complete the entire RSP and EG process. This is where the symmetries were essential. Because these geometric distributions parametrized by $P_{sc}$ and $P_{sc}$, are completely defined by $|\alpha|^2$ and $\xi$, respectively, regardless of the number of qubits and servers we only had two parameters to optimize. Moreover, the third and last parameter that we used in our optimization is $n_o$. This parameter is a cutoff that defines which rounds were a success and it can be tuned in order to meet higher requirements in the final fidelity of the final $s-$qubit state. 

We describe the sampler in pseudocode:

\begin{algorithm}[H]
\caption{Sampler}
\label{alg:sampler}
\begin{algorithmic}[1]
\Require
Total number of rounds $N$, number of servers $M$, number of qubits $s$, mean-photon number for the RSP $|\alpha|^2$, bright state probability for EG $\xi$, and the cutoff $n_o$. 
\Ensure
$F^*(\alpha,\xi,n_o)\geq F_{min}$, otherwise return 0.
\State Calculate $P_{sc}(\alpha)$ and $P_{ss}(\xi)$.
\State Initialize the number of successful preparations $n_{succ}=0$ and the list to collect the number of attempts per trial $n_t = \{\}$.
\For{$i= 1~\text{\textbf{to}}~N$} 
    \State Draw $M$ samples from a geometric distribution with parameter $P_{sc}$. These are denoted $r_i$ for $i=1,\dots,M$.
    \State Draw $s$ samples from a geometric distribution with parameter $P_{ss}$. These are denoted $n_j$ for $i=2,\dots,s-1$.
    \State Sort $r_i$ in increasing order, denote them by $r'_i$, and select the first $s$ of them.	\Comment{This represents the $s$ firsts RSPs}
    \State Calculate the total number of attempts in this round and append it to $n_t$.
    \If{$r'_s-r'_1\leq n_o$ and $n_i \leq n_e, ~\forall i$}\Comment{Conditions to consider a success}  
	\State $n_{succ} \xleftarrow{} n_{succ} +1$ \Comment{The trial is considered a success and increment the counter}
    \Else 
        \State Do nothing. \Comment{The trial is considered a failure}
    \EndIf
\EndFor
\State Estimate the probability of success $p_{succ} = n_{succ}/N$ and the mean number of trials $\left< n_{t} \right>$.
\State return the rate $r(\alpha, \xi, n_o) = p_{succ}/\left< n_{t} \right>$.
\end{algorithmic}
\end{algorithm}

Notice that we simulate the number of attempts it took to complete a RSP $r_i$, where $i$ is the number of the server. And, the number attempts to complete an EG step $n_j$, where $j\geq 2$ denotes the qubit that is teleported towards the server where the first qubit was prepared. The number of attempts can be recast as timetags in a ``wall-time'' approach for the simulations considering that all the RSPs start simultaneously.     

The function $F^*$ in Algorithm \ref{alg:sampler} is the fidelity calculated over the worst case scenario. This means that all steps, RSP and EG, take the largest possible number of attempts ($n'_s-n'_1 = n_o$ and $t_i = n_e$) such that the round has the largest duration, but it is still a success. For instance, for $s=M=2$, this would mean that we calculate the fidelity of a state with $n_2-n_1 = n_o$ and $t_1=n_e$. Using $F^*$ we ensure that any $s-$qubit state prepared will have a fidelity larger than $F_{min}$. Algorithm \ref{alg:sampler} is repeated over various configurations of $(\alpha, \xi, n_o)$, and we choose the configuration that yields the best rates for a given $(s,M)$. 

Notice that due to the nature of this algorithm and the fidelity constraints it is likely that we did not obtain the global maximum of the multiplexed rates. However, our approximation was sufficient to find cases where the classical bound was surpassed.

\end{document}